\documentclass[%
 reprint,
superscriptaddress,
bibnotes,
amsmath,amssymb,
aps,
prl,
floatfix,
]{revtex4-2}

\usepackage{graphicx}
\usepackage{dcolumn}
\usepackage{bm}
\usepackage{chemformula} 

\usepackage{upgreek}

\newcommand{\tu}[1]{\textup{#1}}
\newcommand{\pdvt}[1]{\frac{\partial {#1}}{\partial t}}
\newcommand{\pdvr}[1]{\frac{\partial {#1}}{\partial r}}

\newcommand{\dvt}[1]{\frac{\mathrm{d} {#1}}{\mathrm{d} t}}
\newcommand{\dvtt}[1]{\frac{\mathrm{d}^2{#1}}{\mathrm{d}t^2}}
\newcommand{\RNum}[1]{\uppercase\expandafter{\romannumeral #1\relax}}

\begin{document}


\title{Laser-induced Cavitation for Controlling Crystallization from Solution
}

\author{Nagaraj Nagalingam}
\affiliation{Process \& Energy Department, Delft University of Technology, Leeghwaterstraat 39, 2628 CB Delft, Netherlands}
\author{Aswin Raghunathan}
\affiliation{Process \& Energy Department, Delft University of Technology, Leeghwaterstraat 39, 2628 CB Delft, Netherlands}
\author{Vikram Korede}
\affiliation{Process \& Energy Department, Delft University of Technology, Leeghwaterstraat 39, 2628 CB Delft, Netherlands}
\author{Christian Poelma}
\affiliation{Process \& Energy Department, Delft University of Technology, Leeghwaterstraat 39, 2628 CB Delft, Netherlands}
\author{Carlas S. Smith}
\affiliation{Delft Center for Systems and Control, Delft University of Technology, Mekelweg 2, 2628 CD Delft, Netherlands}
\author{Remco Hartkamp}
\affiliation{Process \& Energy Department, Delft University of Technology, Leeghwaterstraat 39, 2628 CB Delft, Netherlands}
\author{Johan T. Padding}
\affiliation{Process \& Energy Department, Delft University of Technology, Leeghwaterstraat 39, 2628 CB Delft, Netherlands}
\author{Hüseyin Burak Eral}
\email{h.b.eral@tudelft.nl}
\affiliation{Process \& Energy Department, Delft University of Technology, Leeghwaterstraat 39, 2628 CB Delft, Netherlands}

\date{\today}

\begin{abstract}
We demonstrate that a cavitation bubble initiated by a Nd:YAG laser pulse below breakdown threshold induces crystallization from supersaturated aqueous solutions with supersaturation and laser-energy dependent nucleation kinetics. Combining high-speed video microscopy and simulations, we argue that a competition between the dissipation of absorbed laser energy as latent and sensible heat dictates the solvent evaporation rate and creates a momentary supersaturation peak at the vapor-liquid interface. The number and morphology of crystals correlate to the characteristics of the simulated supersaturation peak.
\end{abstract}

\maketitle
Controlling crystallization from solution, which is central to technological applications ranging from nanomaterial synthesis to pharmaceutical manufacturing \cite{Kashchiev2000,Crystallization,doi:10.1126/science.1243022}, is still challenging our understanding of nucleation \cite{PhysRevLett.129.246101,doi:10.1126/sciadv.aav7399,PhysRevLett.126.015704}. Among the strategies proposed to control kinetics and emerging crystal properties \cite{PhysRevLett.107.025504,PhysRevLett.128.166001,10.1021/acs.langmuir.0c00193}, non-photochemical laser-induced nucleation (NPLIN), where one or more unfocused laser pulses trigger accelerated nucleation in supersaturated solutions \cite{10.1103/PhysRevLett.77.3475,10.1016/j.cej.2020.127272,doi:10.1021/ja905232m}, emerged as a promising approach due to its presumed non-chemical nature and ability to influence polymorphic form \cite{PhysRevLett.89.175501,dx.doi.org/10.1039/C7CP03146G}. At the reported laser pulse duration ($\sim$ns), wavelengths (532/1064 nm) and laser intensity ($\sim$MW/cm$^2$), neither the solute nor the solvent have sufficiently strong absorption bands to induce photochemical effects. Several putative mechanistic hypotheses, ranging from molecular phenomena relying on (an)isotropic polarization and isotropic electronic polarizability of solute clusters \cite{doi:10.1021/cg8007415} to microscale phenomena based on impurity heating and consequent cavitation, have been proposed in an attempt to explain the observations \cite{10.1063/1.5079328}. However, the exact mechanism behind NPLIN remains elusive \cite{10.1063/1.5079328}.

Transient micro vapor bubbles can be created in liquid environments with the absorption of laser pulses by dyes \cite{10.1103/PhysRevLett.98.254501} and nanoparticles \cite{10.1103/PhysRevE.85.016319,10.1016/j.expthermflusci.2020.110266}. The impurity heating hypothesis suggests that laser energy absorbed by inherent insoluble impurities (such as nanoparticles) locally evaporates its surrounding solvent - consequently triggering solute nucleation. However, no direct measurements of this hypothesized phenomenon was reported. Most reported NPLIN experiments only quantify the crystallization probability seconds to minutes after laser irradiation \cite{10.1021/acs.cgd.2c01526}. Moreover, the large exposed volumes [$O(\tu{cm}^3)$] and uncertainties in concentration and chemical nature of impurities limit the observation of micron-sized cavitation bubbles within microseconds after laser irradiation. Thus the attempts to test the impurity heating hypothesis using numerical modeling have had limited success due to lack of concomitant experimental data \cite{10.1021/acs.cgd.0c00942}.

In this work, using high-speed microscopy experiments and 1D finite element simulations, we demonstrate that a momentary supersaturation rise surrounding a laser-induced cavitation bubble can trigger crystallization in supersaturated aqueous solutions of potassium chloride (\ch{KCl}). We use a frequency-doubled Nd:YAG pulsed laser with 532\,nm wavelength and 4\,ns pulse duration. Unlike the traditional NPLIN experiments, we focus the laser to fix the location of bubble formation and intentionally dope the solution with a light-absorbing soluble impurity. 3.26\,mg potassium permanganate (\ch{KMnO4}) per 100\,g water, is added to facilitate bubble formation below the optical breakdown threshold via thermocavitation \cite{10.1016/j.expthermflusci.2023.110926} (see S\RNum{1} \cite{SI}). The laser focal spot resembles an impurity being heated up and a consequent cavitation bubble formation, establishing the connection between NPLIN experiments conducted with unfocussed laser and this study. The added \ch{KMnO4} is comparable to the impurities level in traditional NPLIN experiments [$O(10 \,\tu{ppm})$] \cite{10.1021/acs.cgd.7b01277}, and therefore does not alter the solubility of \ch{KCl} (see S\RNum{2} \cite{SI}). Thus, this work differentiates itself from cavitation-induced crystallization experiments via multiphoton absorption using focused ultrashort laser pulse ($\sim$fs) that might involve photochemistry \cite{10.1016/j.jphotochemrev.2022.100530}. Moreover, it captures the size of cavitation bubbles [$O(100\,\upmu$m)] \cite{10.1016/j.expthermflusci.2020.110266} surrounding the nanoparticles for the inferred magnitude of laser energies and impurity sizes in NPLIN experiments \cite{10.1021/acs.cgd.7b01277} (see S\RNum{3} \cite{SI} for calculations). 

We perform experiments to record the size of the vapor bubbles created, the resulting number and morphology of crystal(s) formed, and the cumulative nucleation probability at a fixed time lag. Subsequently, using simulations, we estimate the local temperature, solute concentration, and solute supersaturation surrounding the bubble to complement the experiments. The quantitative agreement between experimental and simulated bubble dynamics validates the proposed model. Leveraging the model, we argue that a competition between the dissipation of absorbed laser energy as latent and sensible heat dictates the instantaneous solvent evaporation rate. A spike in evaporation rate during the cavitation bubble expansion creates a momentary supersaturation peak at the vapor-liquid interface (hereinafter referred to as ``interface''). The experimentally acquired nucleation probabilities, number, and morphology of crystals formed correlate with the characteristics of the short-lived [$O(\upmu\tu{s})$] supersaturation peak surrounding the bubble from simulations. For the first time, we quantitatively correlate the likelihood of crystal formation due to an increase in the solute concentration at the interface through laser-induced bubble formation with no expected photochemical reaction.

\begin{figure}[tbp]
	\centering
	\includegraphics[width=\columnwidth]{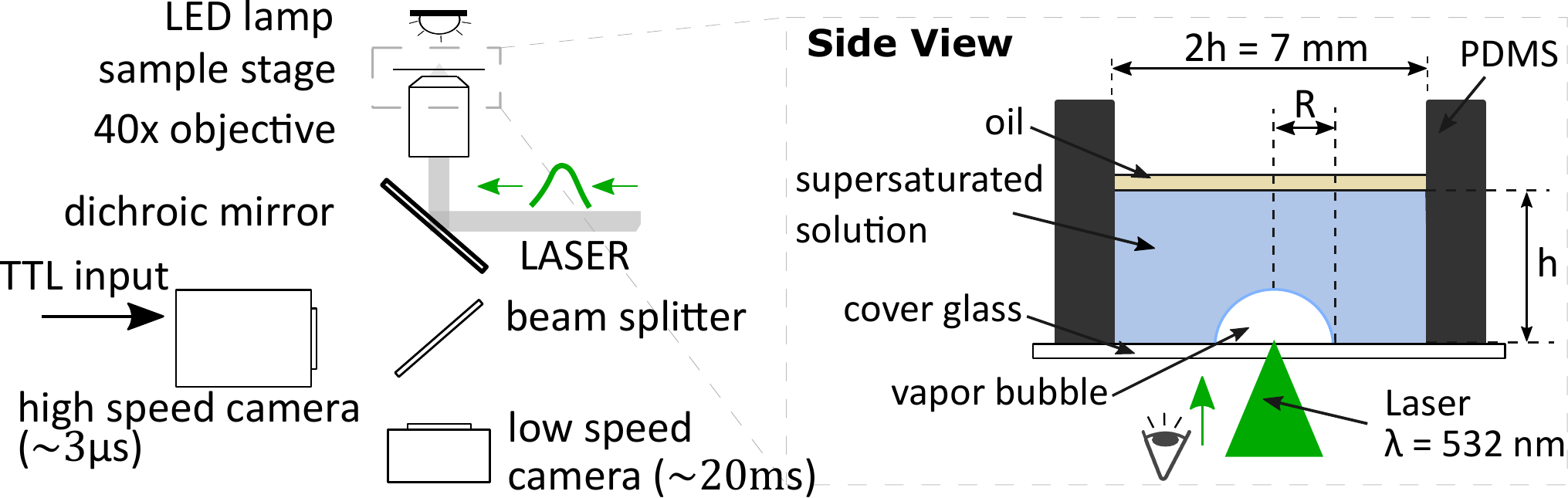}
	\caption{Sketch of the experimental setup to generate a microbubble. The setup construction is detailed in our previous work \cite{10.1016/j.ohx.2023.e00415}. The green arrow indicates the direction of laser pulse.}
	\label{Figure 1}
\end{figure}

In our experiments, KCl solutions with a supersaturation range of $0.999-1.029$ were used (solubility = 35.97\,g/100\,g-\ch{H2O} at $25^{\circ}$C) with no pre-treatment for dissolved gases or filtration. A 40\texttimes\ objective (numerical aperture=0.6) is employed to both focus the laser and image the sample. Fig. \ref{Figure 1} shows the architecture of the inverted microscope which employs two cameras: a high-speed camera operated at 330,000 frames per second (fps) to record the evolution of the bubble size and a low-speed camera operated at 50 fps which records the appearance of crystals. A 1.23\,mm layer of silicone oil (density = $930$\,kg/m$^3$) floating on top of the supersaturated solution prevents evaporation of the solution. The laser is focused to a point within $10\,\upmu$m above the bottom surface (cover glass). The standoff distance to the bottom surface is maintained below 0.05 to prevent surface erosion \cite{10.1016/j.ultsonch.2022.106131}. In addition, all formed hemispherical bubbles in this work have, $h/R_\tu{max}>10$, to prevent the effect of side walls on the bubble dynamics \cite{10.1017/jfm.2016.463}. Thus the cover glass acts as a plane of symmetry for the semi-unbounded fluid surrounding the hemispherical bubble, allowing us to analyze the bubble as spherically unbound - a 3D bubble. Since the negatively buoyant crystals sediment to the bottom, the adapted experimental technique allows in-situ recording of both the bubble and crystal(s). 

\begin{figure}[tbp]
	\centering
	\includegraphics[width=\columnwidth]{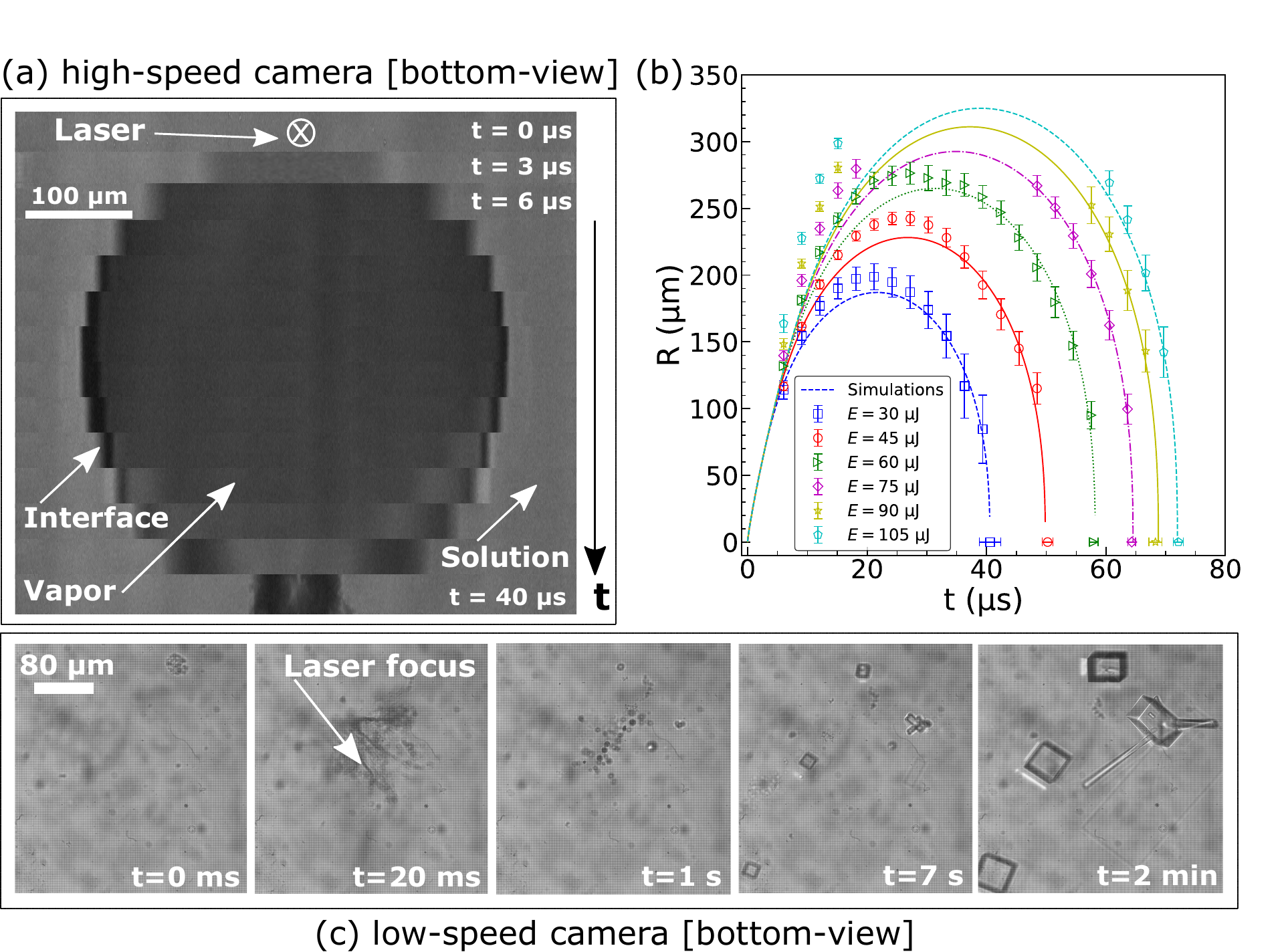}
	\caption{(a) Primary vapor bubble formation using a focused laser pulse of 30 $\upmu$J recorded at 330,000\,fps with a reduced spatial resolution. (b) Dynamic radius of the hemispherical bubble for different laser energies $E$. The error bars represent the standard error on the mean of at least 20 independent trials. A bubble radius beyond $\approx 300\,\upmu$m exceeded the field of view of the camera. The symbols and lines correspond to experiments and simulations, respectively. (c) Secondary bubbles and emergence of crystals after collapse of the primary vapor bubble surrounding the laser focal spot visualized at 50 fps using the low-speed camera. The experiment is for $E=75\,\upmu$J and $S=1.019$.}
	\label{Figure 2}
\end{figure}
Figure \ref{Figure 2}(a) depicts the primary bubble formation, its subsequent expansion, and collapse immediately after laser irradiation. The primary bubble then disintegrates into secondary bubbles followed by the emergence of crystals surrounding the laser focal point (Fig. \ref{Figure 2}(c)). After the primary bubble collapsed, we also observe a complex flow pattern that transports secondary bubbles and crystals. The direction of the resulting flow was observed to be random, consistent with previous observations \cite{10.1103/PhysRevE.80.047301}. Figure \ref{Figure 2}(b) displays a clear increase in the maximum radius ($R_\tu{max}$) and bubble lifetime with the supplied laser energy ($E$). For details on the experimental methodology and validity of the bubble shape, see S\RNum{4} \cite{SI}. 

\begin{figure}[tbp]
	\centering
	\includegraphics[width=\columnwidth]{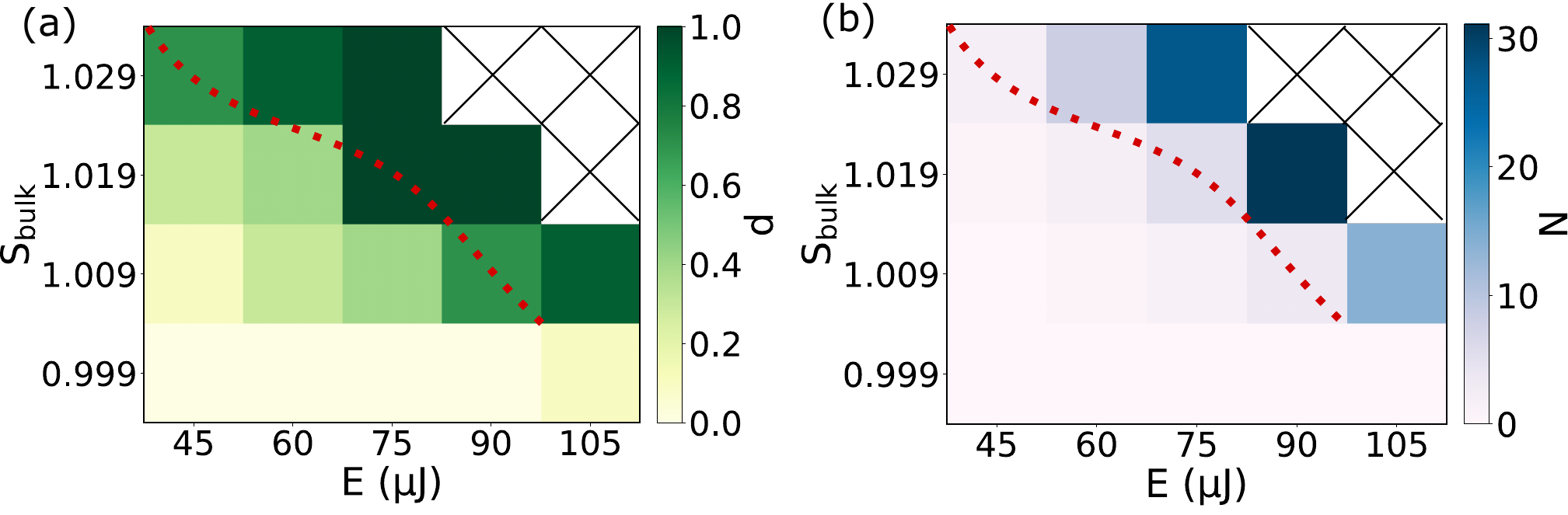}
	\caption{Experimentally observed nucleation statistics: (a) cumulative nucleation probability ($p$) and (b) mean crystal count ($N$), for different laser energies ($E$) and solution supersaturation in the bulk ($S_\tu{bulk}$). The results are for 10 trials, each with a fixed lag of 2 minutes from the time of laser irradiation. The red dotted curve is a guide to the eye representing the threshold where the crystallization probability is $\geq 0.5$. See S\RNum{5} \cite{SI} for morphologies.}
	\label{Figure 3}
\end{figure}
We quantify crystallization by plotting nucleation probability and crystal count for varying laser energy and supersaturation in the bulk, Fig. \ref{Figure 3}. The cumulative nucleation probability ($p$) is defined as the number of trials that resulted in crystal formation two minutes after laser irradiation to the number of trials performed. Overall, the nucleation probability increases with increasing laser energy and solution supersaturation in the bulk ($S_\tu{bulk}$). From Fig. \ref{Figure 3}(a), we observe a minimum threshold laser energy for crystal formation related to $S_\tu{bulk}$ and vice versa, an observation repeatedly reported in NPLIN experiments \cite{10.1021/acs.cgd.2c01526}. We recorded a very low crystallization probability ($p\leq0.1$) for roughly saturated solution ($S_\tu{bulk} = 0.999$) as the lack of supersaturation would inhibit crystal growth. We attribute the non-zero $p$ value to the uncertainty in $S_\tu{bulk}\,[{O}(10^{-3})]$ pertaining to the variation in room temperature ($24.8-26.1\,^{\circ}$C). No experiment was performed beyond $S_\tu{bulk} = 1.029$ as it was difficult to keep the solution stable during handling. In Fig. \ref{Figure 3}(b), similar to the nucleation probability, we see an increase in the number of crystals formed ($N$) with both laser energy and bulk supersaturation above the minimum laser intensity threshold. We predominantly observed cubic crystals with the probability of finding a rectangle or needle-like crystal increasing with $E$ and $S_\tu{bulk}$ (see S\RNum{5} \cite{SI}). This observed change in morphology aligns with previous observations \cite{Kardum,10.1021/acs.jpcc.9b11651}, deduced using limited solvent availability per nuclei. In our experiments, we cannot measure local fluid properties surrounding the bubble, such as temperature and solute concentration, due to the small length and time scales involved. Therefore, we employ numerical simulations to calculate temporal and spatial values of these variables while the experimentally measured bubble radii and crystal count are used to validate the fluid flow and local supersaturation, respectively. 

In numerical simulations, we solve for combined momentum, heat, and solute transport. For each phenomenon, the governing equations for an unbound 3D sphere are used due to the plane of symmetry offered by the cover glass. We employ the Rayleigh-Plesset equation \cite{10.1146/annurev.fl.09.010177.001045} to solve for the momentum surrounding the bubble,
\begin{equation}
    R\dvtt{R} + \frac{3}{2} \left(\dvt{R}\right)^2  = \frac{1}{\rho_\tu{L}} \left(p_\tu{V}-p_\infty -\frac{2\sigma}{R} - \frac{4\mu}{R}\dvt{R}\right), 
    \label{rp equation}
\end{equation}
where $\rho_\tu{L}=1175$\,kg/m$^3$ is the solution density, $p_\infty=1.013$\,bar is the ambient pressure and $p_\tu{V}$ is the pressure within the bubble, $\sigma$ is the surface tension, $\mu$ is the dynamic viscosity of the solution and $R$ the distance of the interface from the laser focal point. The spherically symmetric heat dissipation surrounding the bubble is modeled using,
\begin{equation}
    \pdvt{T} + \frac{R^2}{r^2}\dvt{R} \pdvr{T} = \frac{1}{r^2} \frac{\partial}{\partial r}\left(r^2 \alpha \pdvr{T}\right),
    \label{heat diffusion}
\end{equation}
in which $T$ is the temperature, $\alpha$ is the thermal diffusivity of the solution and $r$ ($>R$) the radial position from the bubble center. For solute transport, we use an analogous equation to Eq. (\ref{heat diffusion}) by substituting $T$ with $C^*$, the solute concentration in kg/kg of the solution, and $\alpha$ with $D$, the mass diffusivity of the solute.

For simplicity, we assume the bubble to be a lumped system with an energy balance given by
\begin{equation}
    \dvt{(m_\tu{V} c_{p\tu{V}} T_\tu{V})} + \dvt{m_\tu{V}} H_\tu{L} = A_\tu{V} k \left(\pdvr{T}\right)_{r=R},
    \label{heat conversion}
\end{equation}
where $m_\tu{V}$, $A_\tu{V}$, and $c_{p\tu{V}}$ are the mass, surface area, and specific heat capacity of the vapor bubble, respectively. $H_\tu{L}$ is the latent heat of vaporization and $k$ the thermal conductivity of the solution. At the interface, we enforce the boundary condition $T_\tu{V} = T|_{r=R}$ at all times, where $T_\tu{V}$ is the bubble temperature. The change in mass of the bubble is estimated using the corrected Hertz-Knudsen equation \cite{10.1016/0021-9797(92)90205-Z}, $\mathrm{d}m_\tu{V}/\mathrm{d}t = -(16 A_\tu{V} / 9 \sqrt{2 \pi \tu{R}_\tu{g} T}) [p_\tu{V} - p_\tu{sat}(T|_{r=R})]$, where $\tu{R}_\tu{g}$ is the specific gas constant for water vapor and $p_\tu{sat}(T|_{r=R})$ the saturation pressure of the solution at the interface. The $p_\tu{V}$ is estimated using the ideal gas law, $p_\tu{V} V_\tu{V}=m_\tu{V}\tu{R}_\tu{g}T_\tu{V}$, where $V_\tu{V}$ is the bubble volume.

At $t=0$ the bubble is assumed to be saturated with zero interface velocity, and the surrounding solute concentration is assumed to be same as in the bulk. System energy is imposed by initializing a thermal boundary layer profile surrounding the bubble. The initial temperature distribution is $T(\xi) = T_\infty + (T_\tu{V}-T_\infty) \exp[-(\xi / \delta_\tu{T})^{25}]$, where $T_\infty$ is the ambient temperature, $\delta_\tu{T}$ is the thermal boundary layer thickness and $\xi = r-R$, is the radial distance from the interface. A high exponent of 25 is used to approximate a step function, while still being smooth enough to avoid numerical instabilities near $\xi \approx \delta_\tu{T}$. The thermal energy supplied in the simulation is transformed into latent heat (vapor), sensible heat (vapor and liquid), and kinetic energy of the solution. Thus, the control parameter in simulation, $\delta_\tu{T}$, characterizes the energy available for a bubble to grow. The initial bubble temperature is chosen to be 650\,K - the spinodal temperature \cite{10.1007/s10953-009-9417-0}, with the radius $0.5\,\upmu$m calculated using theoretical laser spot dimensions. For details on the numerical model and parameter values solved, refer S\RNum{6} \cite{SI}.

\begin{figure*}
    \centering
    \includegraphics[width=\linewidth]{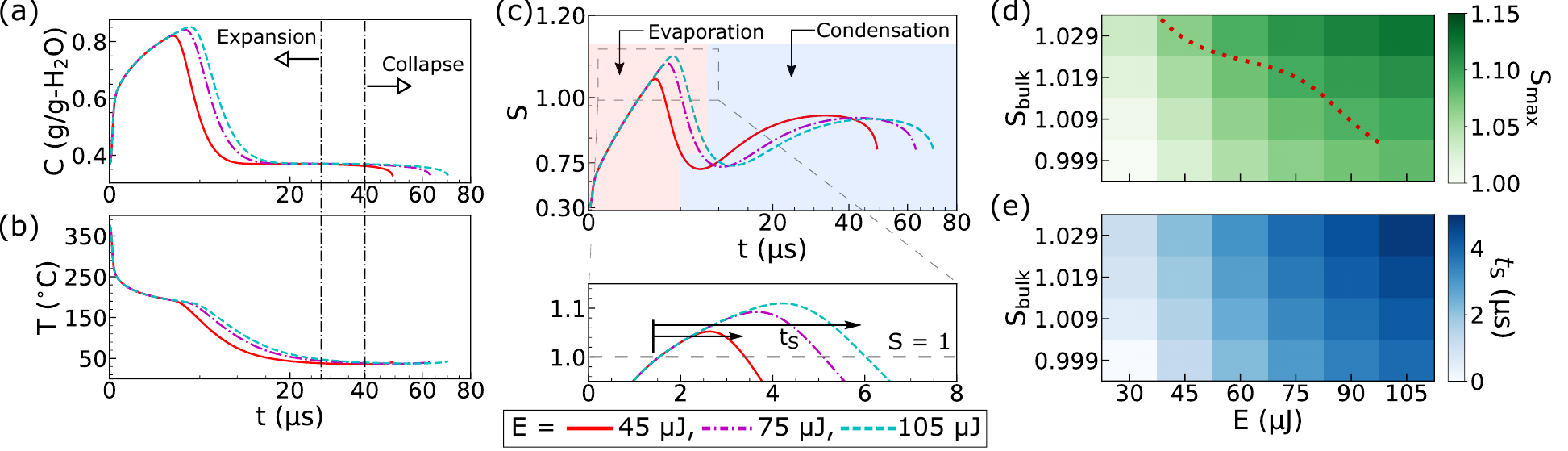}
    \caption{(a,b) Simulated temporal change in solute concentration ($C$) and temperature ($T$) at the interface for $S_\tu{bulk} = 1.019$ (at $25^{\circ}$C). Since the laser pulse duration (4\,ns) is negligible compared to the time scale of the phenomena ($\upmu$s), we consider the energy transfer from the laser ($E$) to the solution to be instantaneous at $t$=0 ($x$-axis is scaled quadratically). (c) The supersaturation ratio calculated using the concentration and temperature plotted in (a) and (b), respectively. The $x$-axis scale is quadratic, while the $y$-axis scale is cubic. $t_\tu{S}$ represents the time period for which $S>1$. (d) The simulated maximum $S$ values obtained for all the conditions in this work, similar to the examples from (c). The red dotted curve is the guide to the eye from Fig. \ref{Figure 3}, representing the crystallization probability $\geq 0.5$ in experiments. (e) The time period for which the simulated $S>1$, similar to the examples from (c).}
    \label{Figure 4}
\end{figure*}
Figure \ref{Figure 2}(b) shows the numerically obtained bubble size for $\delta_\tu{T} =$ 21, 25.5, 29.5, 32.5, 34.5 and 36 $\upmu$m, corresponding to the increasing laser energy values from experiments (see S\RNum{6} \cite{SI} for calculations). The deviation between experiments and simulations in $R$ for higher energies ($E\geq 90\,\upmu$J) can be attributed to non-linear absorption \cite{10.1109/3.777215,10.1166/jnn.2016.11842} with possible plasma formation. The plasma can initiate high pressures, leading to higher interface velocities \cite{10.1121/1.415878}. The probability of bubble incidence with and without \ch{KMnO4} in water was investigated for non-linear absorption, which supports the reasoning made for deviations in $R$ (see S\RNum{1} \cite{SI}). Moreover, the increase in interface velocities will only enhance the solvent accumulation at the interface supporting our hypothesis (see S\RNum{6} \cite{SI}).

To get insight into the crystal formation surrounding the bubble, we look at the factors affecting the solute supersaturation using simulation. Figure \ref{Figure 4}(a,b) shows the temporal evolution of the solute concentration and temperature at the dynamic interface for three different laser energies at fixed bulk supersaturation. Initially, the temperature drops abruptly, in conjunction with a steep rise in concentration due to high evaporation rates, ${O}$(100\,kg/(m$^2$s)). Then, the decrease in temperature is more gradual, while the decrease in concentration is steep. The drop in temperature can be attributed to heat diffusion away from the interface and advection resulting from bubble dynamics. Similarly, for the solute, there is dilution occurring at the interface due to condensation of the vapor in addition to diffusion and advection. During the latter half of the bubble lifetime, the concentration and temperature have minimal change due to lower driving potentials and short time range, ${O}$(10 $\upmu$s). The temperatures during bubble collapse estimated from the simulations are in good agreement with the empirical calculations from literature (see S\RNum{4} \cite{SI}).

 Figure \ref{Figure 4}(c) shows the temporal supersaturation at the interface calculated using profiles given in Fig. \ref{Figure 4}(a,b). We observe a peak in the local supersaturation ratio when the bubble is rapidly expanding, after which the supersaturation decreases and the interface stays undersaturated ($S<1$) within the bubble lifetime. This observation of a momentarily supersaturated state ($S>1$), highlighted in the close-up in Fig. \ref{Figure 4}(c), is a favorable condition for crystal nucleation. Moreover, both the peak supersaturation ($S_\mathrm{max}$) and the time during which the interface remains supersaturated ($t_\tu{S}$) increase with increasing $E$. In the above analysis, we only look at the interface since heat diffuses faster than the solute and thus the maximum supersaturation ratio can exist only in the region closest to the bubble, i.e., at the interface. The estimated supersaturated layer thickness increases with $E$ and is ${O}$(10\,nm), consistent with literature \cite{10.1021/acs.cgd.0c00942}. However, this supersaturation ratio at the interface is dynamic and is quantified only when the bubble exists. The induced flow and resulting temperature and solute distribution surrounding the laser focal point after the bubble collapses are complex and outside the scope of this work. The simulated trends observed in Fig. \ref{Figure 4}(d,e) agree well with the presented experimental results in Fig. \ref{Figure 3}.

\begin{figure}[tbp]
	\centering
	\includegraphics[width=\columnwidth]{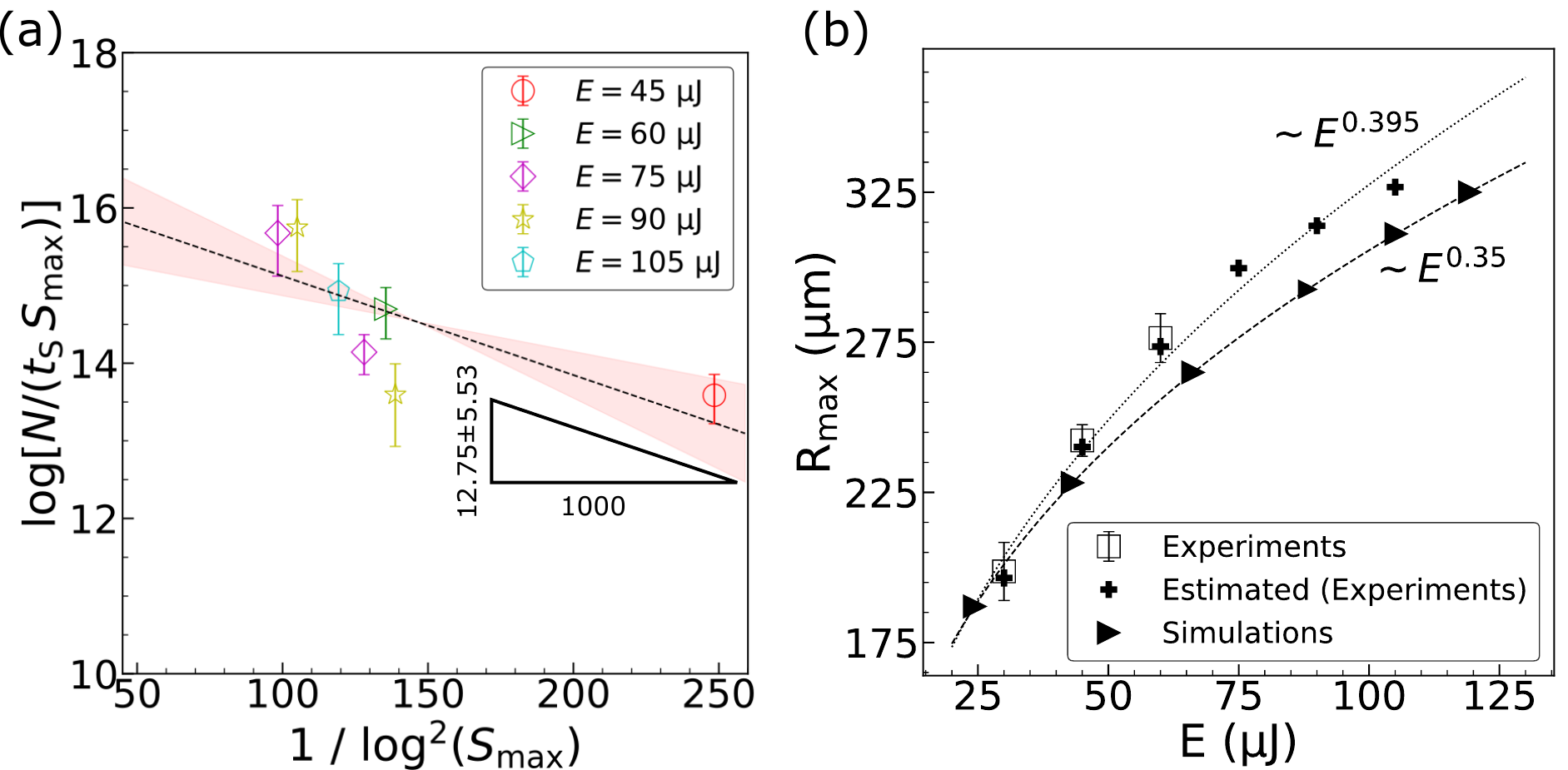}
	\caption{(a) Estimate of nucleation rate $J$ against simulated peak supersaturation ($S_\tu{max}$). $J \propto N/t_\mathrm{S}$, where $N$ is the mean crystal count from experiments and $t_\mathrm{S}$ is the time for which $S>1$ in simulations. (b) Maximum vapor bubble radius plotted against the energy supplied. The dotted and dashed lines represent the power law fit for the data from experiments and simulations, respectively. Error bars represent the standard error on the mean.}
	\label{Figure 5}
\end{figure}
Subsequently, we correlate the simulated crystallization parameters, $S_\mathrm{max}$ and $t_\tu{S}$, with the experimentally acquired parameter, $N$ (Fig. \ref{Figure 3}b). The nucleation rate (the number of nuclei formed per unit time per unit volume) can be expressed as \cite{10.1002/crat.200310070}, 
\begin{equation}
    J \propto S\exp[-16\pi v^2\gamma^3 / (3k_\tu{B}^3 T^3\log^2(S))].
    \label{J eqn}
\end{equation}
where $\gamma$ is the solute-solution interfacial tension, $k_\tu{B}$ is Boltzmann's constant and $v$ the molecular volume. We relate $J \propto N/t_\tu{S}$. Since the size of the bubbles for the time region where $S>1$ are almost the same within the range of energies used, we leave out the shell volume surrounding the interface in the proportionality for $J$. Using the slope from Fig. \ref{Figure 5}(a), we estimate $\gamma$ in Eq. (\ref{J eqn}) to be $3.7\substack{+0.47 \\ -0.65}$\,mJ/m$^2$ (at $\approx 185-191\, ^{\circ}$C). This value, when calculated for $25\, ^{\circ}$C (3.51\,mJ/m$^2$), is within the reported values of 2.19-5.283\,mJ/m$^2$ for NPLIN \cite{10.1021/cg5004319,10.1021/cg8007415,10.1021/cg300750c} (see S\RNum{7} \cite{SI} for calculation). Note that the elevated temperature is also a favorable condition for crystal nucleation in addition to supersaturation (Eq. \ref{J eqn}). 

Figure \ref{Figure 5}(b) is an equivalent representation of Fig. \ref{Figure 2}(b), showing the dependence of maximum bubble size ($R_\tu{max}$) for varying supplied energies. The estimate of the $R_\tu{max}$ in experiments was made using the bubble lifetime \cite{10.1017/jfm.2015.183,10.1529/biophysj.105.079921} (see S\RNum{8} \cite{SI}). The closely matching trends between experiments and simulations support the reliability of the boundary conditions and assumptions employed in the simulation.

In summary, we have shown that primary nucleation in supersaturated aqueous \ch{KCl} solution can be triggered by thermocavitation induced by a single Nd:YAG laser pulse below the optical breakdown threshold. The nucleation probability as well as the number and morphology of crystals formed depends on bulk supersaturation and laser energy used. Combining high-speed microscopy experiments and finite element simulations, we propose a nucleation mechanism based on the solute accumulation at the interface due to solvent evaporation into the growing bubble. Simulations reveal a momentary spike in supersaturation with a lifetime [$O(\upmu \tu{s})$] proportional to the bulk supersaturation and the supplied laser energy to facilitate nucleation.

The proposed mechanism is distinct from other speculated routes to crystal nucleation in laser-induced cavitation experiments, for example, due to photochemistry \cite{10.1021/cg049604v} and shock waves \cite{10.1103/PhysRevLett.99.045701}. The intentional addition of \ch{KMnO4} enabled bubble formation via thermocavitation avoiding photochemistry - that otherwise might exist due to plasma in cavitation via optical breakdown. Furthermore, our calculations reveal the lengthscale of shockwave influencing crystallization [$O(10\,\upmu$m)] matches the thermal boundary layer thickness surrounding the bubble (see S\RNum{8} \cite{SI}). Therefore, we expect no formation of crystals
due to shockwaves because of the lower supersaturation ratio associated with higher temperatures. Thus, the proposed mechanism, verified by combining experiments and simulations, may shed light on the discussion of the working mechanism(s) behind NPLIN and sonocrystallization via cavitation \cite{10.1021/acs.cgd.8b00547,10.1103/PhysRevLett.124.034501}. 

\begin{acknowledgments}
This work was funded by LightX project under NWO Open Technology Programme (project number 16714). We thank Dr. D. Irimia and Ing. E.F.J. Overmars for supporting the experiments, and S\'ara B\'anovsk\'a for supporting the solubility tests. A special thanks to Dr. H.J.M. Kramer, Dr. A.E.D.M. van der Heijden and members of the LightX user committee for the productive discussions.
\end{acknowledgments}

\bibliographystyle{apsrev4-2}
\nocite{*}
\bibliography{refs}

\end{document}



\title{Supplementary Information:\\Laser Induced Cavitation for Controlling Crystallization from Solution
}

\author{Nagaraj Nagalingam}
\affiliation{Process \& Energy Department, Delft University of Technology, Leeghwaterstraat 39, 2628 CB Delft, Netherlands}
\author{Aswin Raghunathan}
\affiliation{Process \& Energy Department, Delft University of Technology, Leeghwaterstraat 39, 2628 CB Delft, Netherlands}
\author{Vikram Korede}
\affiliation{Process \& Energy Department, Delft University of Technology, Leeghwaterstraat 39, 2628 CB Delft, Netherlands}
\author{Christian Poelma}
\affiliation{Process \& Energy Department, Delft University of Technology, Leeghwaterstraat 39, 2628 CB Delft, Netherlands}
\author{Carlas S. Smith}
\affiliation{Delft Center for Systems and Control, Delft University of Technology, Mekelweg 2, 2628 CD Delft, Netherlands}
\author{Remco Hartkamp}
\affiliation{Process \& Energy Department, Delft University of Technology, Leeghwaterstraat 39, 2628 CB Delft, Netherlands}
\author{Johan T. Padding}
\affiliation{Process \& Energy Department, Delft University of Technology, Leeghwaterstraat 39, 2628 CB Delft, Netherlands}
\author{Huseyin Burak Eral}
\email{h.b.eral@tudelft.nl}
\affiliation{Process \& Energy Department, Delft University of Technology, Leeghwaterstraat 39, 2628 CB Delft, Netherlands}

\date{\today}

\maketitle

\tableofcontents

\section{Bubble incidence in experiments}
\label{Bubble incidence in experiments}
\begin{figure}[htbp]
	\centering
	\includegraphics[width=0.7\columnwidth]{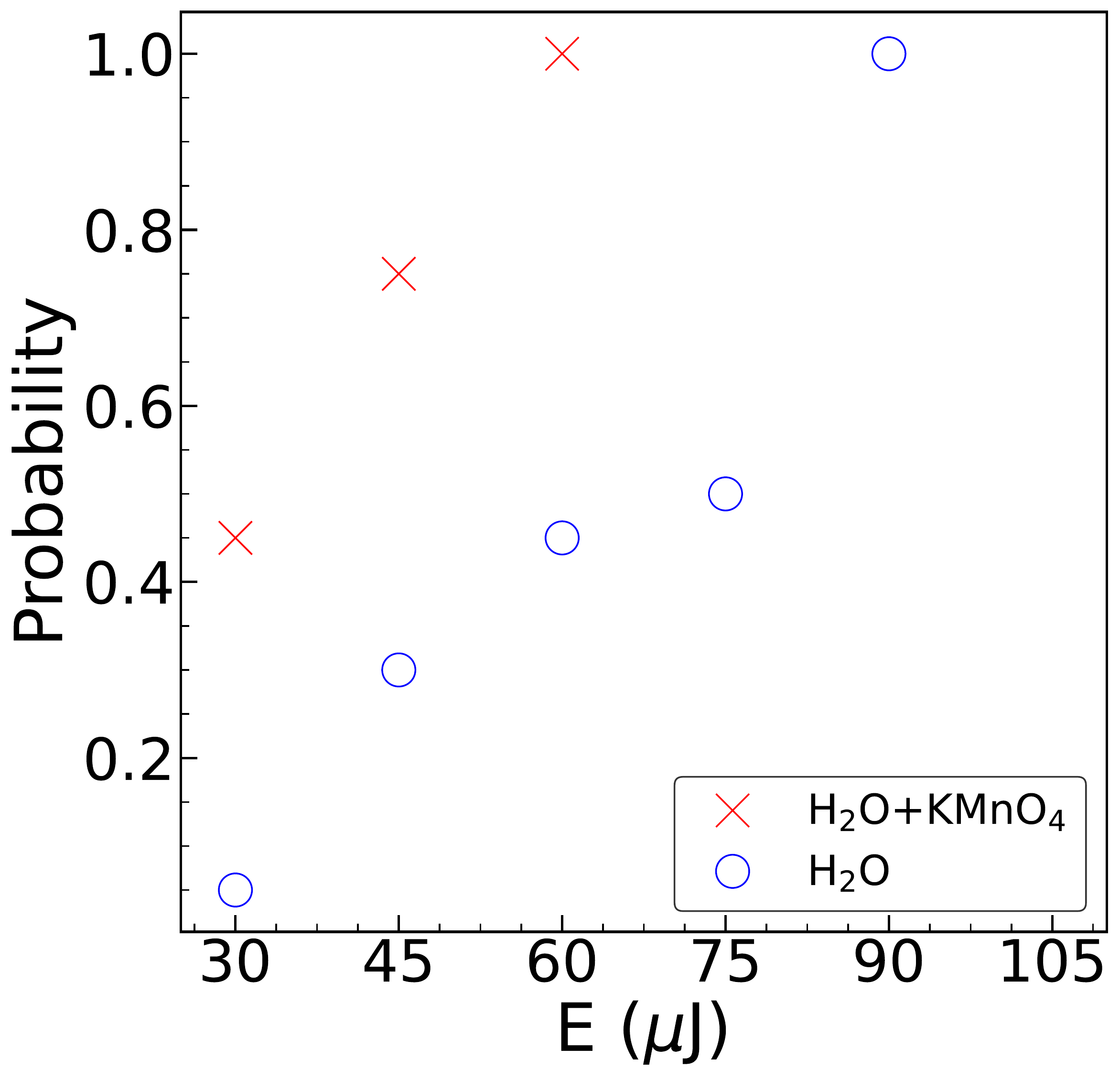}
	\caption{Probability of bubble incidence in experiments. The results are averaged over 20 trials.}
	\label{plasma incidence}
\end{figure}

Figure \ref{plasma incidence} shows the probability of bubble incidence observed in our experiments with ultrapure water (18.2\,M$\Omega$\,cm, ELGA Purelab) and aqueous \ch{KMnO4} (3.26 mg per 100 g of water). Since water has no absorbance to 532\,nm light we attribute the observed bubble formation to non-linear absorption with the possibility of plasma formation.     
\begin{figure}[htbp]
	\centering
	\includegraphics[width=0.7\columnwidth]{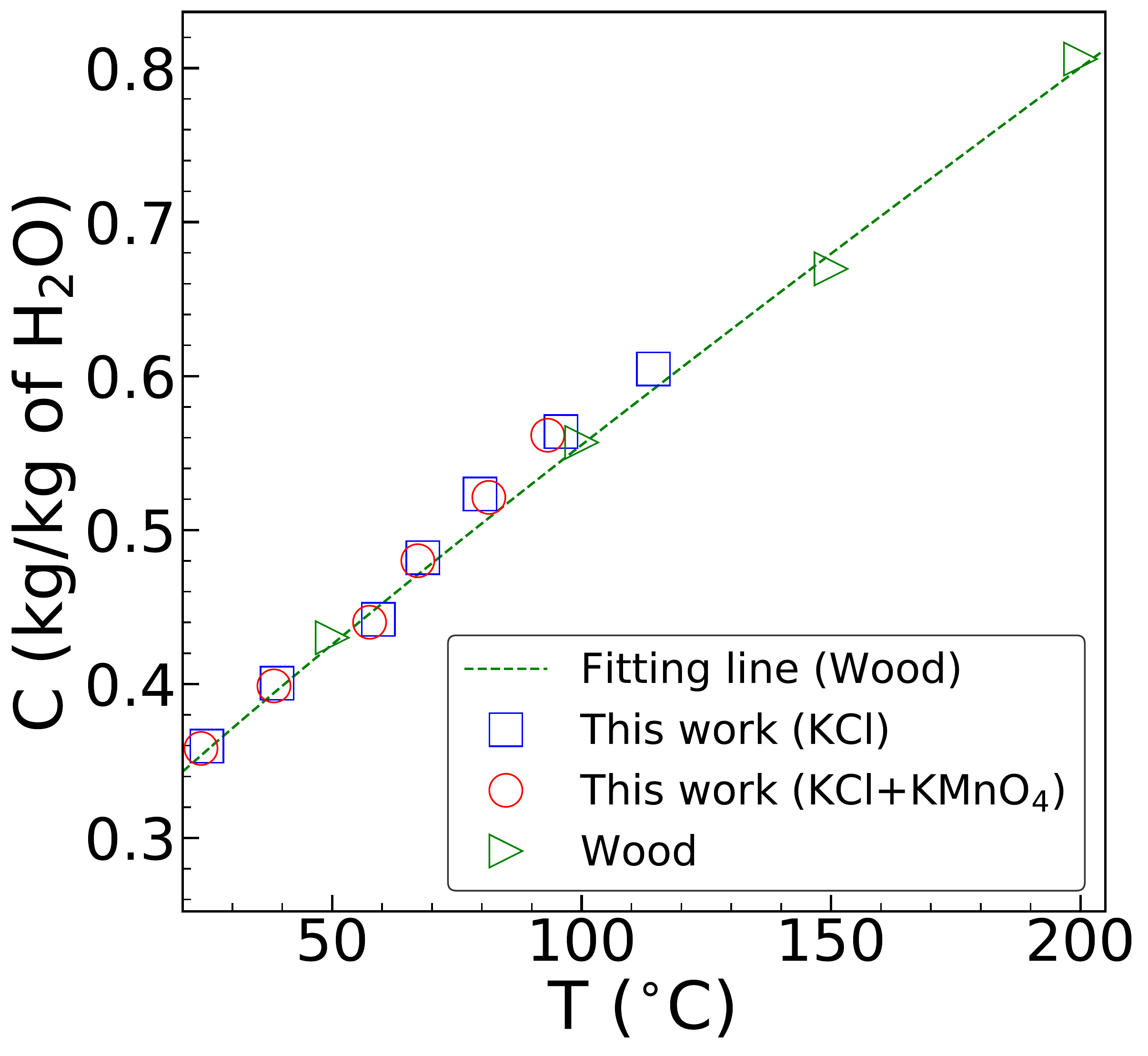}
	\caption{\ch{KCl} solubility estimated using turbidity measurements. Each data point corresponds to an average of 3 trials (except T = 115\,$^{\circ}$C, which contains one). The experimental results from this work are compared with those reported by Wood \cite{10.1016/0016-7037(76)90156-3}.}
	\label{solubility}
\end{figure}

\section{Solution preparation and Solubility}
Ultrapure water (18.2\,M$\Omega$\,cm, ELGA Purelab) was used to prepare the supersaturated \ch{KCl} (P9541, Sigma-Aldrich) solutions in a 8 \,ml vial (SF8, BGB). 30\,$\upmu$l \ch{KMnO4} (223468, Sigma-Aldrich) of 765.92\,mg per 100\,g of H$_{2}$O is added to 7\,ml \ch{KCl} solution to improve its light absorbance at 532\,nm. After preparation, the solution is placed in the oven maintained at 65$^{\circ}$C overnight for complete dissolution.

The solubility measurements were conducted using an equipment (Crystalline, Technobis Crystallization Systems) that measures the turbidity (transmissivity) of the solution to estimate the solubility line (clear point). The experiments were performed using 8\,ml HPLC vials stirred at 700\,rpm subjected to a heating rate of 0.1$^{\circ}$C/min. The clear point is estimated as when the solution reaches 100\% transmission. The experimentally measured solubility from this work is presented in Fig. \ref{solubility}. The solubility of \ch{KCl} (without \ch{KMnO4}) was measured to be 35.97\,g/100\,g-\ch{H2O} at $25^{\circ}$C, and is the value used for estimating supersaturation in the bulk.

The solubility data from Wood \cite{10.1016/0016-7037(76)90156-3} is best fit using the relation $(280.54+3.97\,(T-273)^{0.92})\times10^{-3}$, where $T$ is the temperature in Kelvin.

\section{Predicted Bubble Sizes in NPLIN}
\label{Predicted Bubble sizes in NPLIN}
In relation to the impurity heating mechanism proposed for NPLIN, we predict the size of microbubbles hypothesized to appear around heated nano-impurities using an empirical relation from literature \cite{10.1016/j.expthermflusci.2020.110266} employing Mie theory \cite{Mie_theory}. Comparing the predicted bubble size to those observed in this work provides a sanity check.

\begin{figure}[htbp]
	\centering
	\includegraphics[width=0.8\columnwidth]{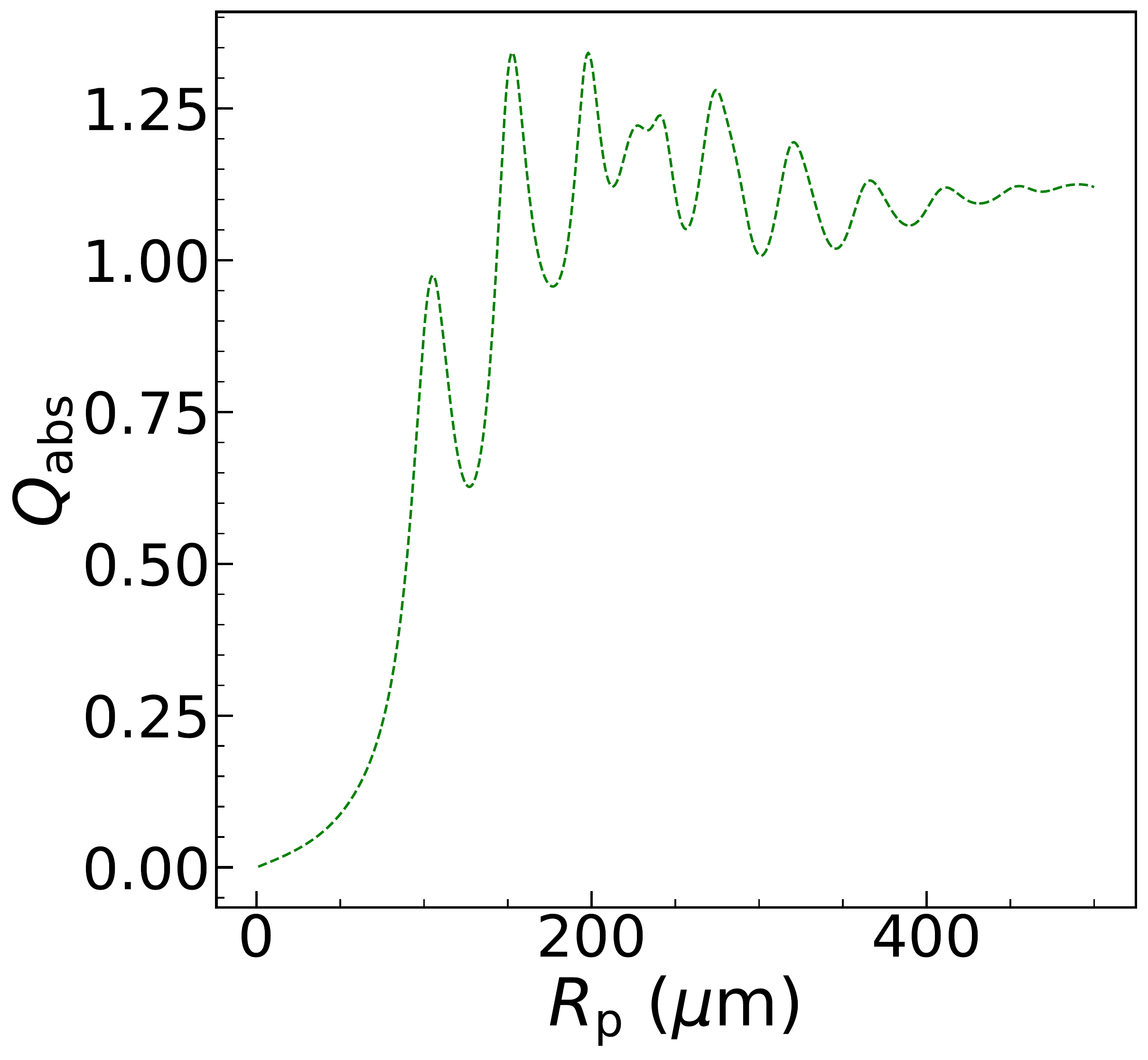}
	\caption{Absorption efficiency estimated using Mie theory \cite{Mie_theory}. The value for the complex refractive index is $2.3456 + i\, 0.0926$ (at 532 nm) \cite{querry1985optical}.}
	\label{Qabs}
\end{figure}

The elemental analysis of \ch{KCl} from the manufacturer indicates the element iron (Fe) as the major impurity ($\leq 3$\,ppm). The relationship between the vapor bubble radius and the insoluble iron nanoparticle (\ch{Fe3O4} - one of the most common states of \ch{Fe}) radius is empirically approximated as $R_\tu{max} \propto (Q_\tu{abs} R_\tu{p}^2 I_0)^{(1/3)}$ \cite{10.1016/j.expthermflusci.2020.110266}. $R_\tu{p}$ is the radius of the nanoparticle, $I_0$ is the laser intensity and $Q_\tu{abs}$ the absorption efficiency calculated using Mie theory \cite{Mie_theory} (see Fig. \ref{Qabs}). The experiments by Kacker \etal\cite{10.1021/acs.cgd.7b01277}, used a filter with pore size of 0.45 $\upmu$m and reported a minimum laser energy threshold of $\approx 5$ MW/cm$^2$ \cite{10.1021/cg8007415}, for \ch{KCl} crystallization with $S_\tu{bulk}=1.048$. In our calculations, we therefore use a nanoparticle radius of 225 nm and estimate the vapor bubble radius to be 170 $\upmu$m, when compared to a 141 $\upmu$m radius bubble formed using 50 nm nanoparticles and 790 MW/cm$^2$ laser intensity \cite{10.1016/j.expthermflusci.2020.110266}. By roughly extrapolating our experimental data to $S_\tu{bulk}=1.048$, the estimated bubble size should be a sufficient condition for crystallization. 

\section{Experimental methodology and validation}
\label{Experimental methodology and validation}
\noindent \textbf{Sample preparation}\\
\indent The microfluidic device side walls are fabricated using polydimethylsiloxane (PDMS, Sylgard\texttrademark\ 184, Dow) and a cover glass with a thickness of $0.13-0.16$ mm is attached to the bottom side. The device is hydrophobized with trichloro(1H,1H,2H,2H-perfluoroctyl)-silane to prevent PDMS from triggering heterogeneous nucleation. During the experiments, before transferring the solution to the microfluidic device, we remove the solution from the oven (at 50$^{\circ}$C) and stir it at 1000 rpm for 1 min using a magnetic stirrer over a hotplate maintained at 50$^{\circ}$C. The solution is transferred using a micro-pipette while it is still warm to avoid spontaneous nucleation while handling. Immediately we pipette silicone oil (378321, Sigma-Aldrich) over the solution to avoid evaporation. We place the device on the microscopic travel stage and allow it to cool down for approximately 7 minutes to let the solution reach room temperature (25$^{\circ}$C). Prior to shooting the laser, the bottom of the microfluidic device is scanned thoroughly to ensure the absence of any crystals. The entire scanning takes around 5 minutes. Following this, the laser is focused to a location within 10 $\upmu$m (least count of the travel stage) above the bottom surface. The solution impurities settled on the bottom are used to estimate the position of the bottom surface. In benchmark experiments, the location of the laser focus relative to the surface is validated by intentionally damaging the substrate by moving it close to the laser focus.

After each trial, the microfluidic device was cleaned with q-tips and rinsed well with water and methanol. They were then re-hydrophobized before the next trial to ensure a uniform surface layer within the device. Each device was reused (with re-hydrophobization) for a maximum of two times.\\

\noindent \textbf{Optics}\\
\indent For imaging, we use a red LED light ($\lambda_\tu{LED}=625$ nm) and an infinity-corrected 40x objective (LUCPLFLN40X, Olympus) with 0.6 numerical aperture (\textit{NA}) and 4.5 mm focal length ($f$). The resolution of the objective is $\lambda_\tu{LED}$/(2\,\textit{NA}) = 0.52 $\upmu$m. A red LED was chosen since the dichroic filter (MD568, Thorlabs) used in the setup has a transmission band of $580-650$ nm and in addition, the high-speed camera (FASTCAM NOVA S16, Photron) has its maximum spectral response at this particular wavelength. The resolution of the image was measured using a test target (R1L1S1N, Thorlabs), and was found to be 0.5076 $\upmu$m/px (for the high-speed camera) and  0.2016 $\upmu$m/px (for the low-speed camera).  

We use a pulsed Nd:YAG laser (Nano L 50:50 PIV, Litron) with 532 nm wavelength ($\lambda$), 4 ns pulse duration ($t_\tu{p}$) and 4 mm beam diameter ($D_\tu{L}$). The estimated laser spot size diameter, $2w_0 = 0.762$ $\upmu$m, is calculated using the expression $4M^2\lambda f / (\pi D_\tu{L})$, where $\lambda$ is the laser wavelength and the beam quality parameter $M^2=1$ (assuming a perfect Gaussian profile). The depth of field (DOF) of the focused laser beam is $2\pi w_0^2 / (M^2 \lambda) = 1.71$ $\upmu$m.\\

\noindent \textbf{Does the bubble stay hemispherical ?}

As our experimental setup does not allow imaging the bubble from side view, we can not quantify how much the bubble deviates from the hemispherical shape. The bubble nucleates close to the bottom surface ($<10\,\upmu$m) because this is where we focus the laser. After nucleation, the bubble growth is fast and inertia-driven. To test the validity of the hemispherical assumption, we analyze two potential disturbances that can affect the bubble shape, without which the overall appearance of the shape of the bubble would be of a hemispherical cap during the entire growth and collapse process \cite{Supponen}. The arguments are as follows:\\
$\bullet$ \textit{Viscous boundary layer over cover glass}: In reality, there may be some degree of deformation at the vapor-liquid interface near the cover glass (bottom surface), but because of the inertial character of the growth-collapse of the bubble, we expect any wall-tension induced deformation to remain relatively localized (relative to the size of the bubble) close to the cover glass. For illustration, we choose the experiment with largest bubble lifetime ($t_\mathrm{osc}=72\,\upmu$s, refer Fig.\,5 in main text) as the length scale for the diffusive growth of the viscous boundary layer is proportional to $t_\mathrm{osc}$ following the relation $\sqrt{(1.72^2\mu t_\mathrm{osc}/\rho_\tu{L})}$ \cite{batchelor1967}. Where $\mu$ and $\rho_\tu{L}$ are the solution's dynamic viscosity and density respectively. The calculated length scale is $13.4\,\upmu$m. Therefore the deformation should be relatively localized when compared to the corresponding bubble size ($R_\mathrm{max}=325\,\upmu$m, refer Fig.\,5 in main text), supporting the hemispherical bubble assumption.\\
$\bullet$ \textit{Buoyancy}: We quantify rather a bubble can become spherical over time as it can be driven away from the cover glass due to buoyancy. We therefore calculate the terminal velocity ($v_\mathrm{rise}$) in the presence of buoyancy force assuming the bubble is spherical, $v_\mathrm{rise} = 2R_\mathrm{max}^2 (\rho_\mathrm{L} - \rho_\mathrm{V}) g / (9 \mu)$. Where $g$ is the acceleration due to gravity and the density of vapor, $\rho_\tu{V}\ll\rho_\tu{L}$. Taking $R_\mathrm{max}=325\,\upmu$m (estimated maximum bubble size in this work), we calculate the $v_\mathrm{rise}$ to be 0.26\,m/s. Therefore for the corresponding bubble lifetime $t_\mathrm{osc}=72\,\upmu$s, the length scale of bubble rise is $18.72\,\upmu$m (5.76\% of $R_\tu{max}$), supporting the hemispherical bubble assumption.\\

\noindent \textbf{Can the hemispherical bubble in experiments be analyzed as a sphere ?}

We verify that the bubbles in our experiments behave similar to a sphere due to the plane of symmetry offered by the cover glass of the microfluidic device (refer Fig.\,1 in main text). We validate this claim by comparing the lifetime of the hemispherical bubbles against the theory for spherical bubbles using Rayleigh collapse time \cite{10.1080/14786440808635681}.

Figure 2(b) (in the main text) displays a clear increase in the maximum radius ($R_\tu{max}$) and lifetime of the bubble ($t_\tu{osc}$) with the supplied laser energy ($E$). Theoretically, we can compare the dynamics of our hemispherical bubble to an unbound spherical bubble using the time that the bubble takes to collapse from its maximum radius to zero, known as the Rayleigh collapse time ($t_\tu{col}$). For an unbound spherical bubble \cite{10.1080/14786440808635681},
\begin{equation}
    t_\tu{col}/R_\tu{max} = 0.91468 \left[\rho_\tu{L}/(p_\infty-p_\tu{V})\right]^{1/2},
    \label{t_osc}
\end{equation}
where $\rho_\tu{L}=1175$ kg/m$^3$ is the density of the solution, $p_\infty=1.013$ bar is the ambient pressure and $p_\tu{V}$ is the vapor pressure within the bubble. If we assume the bubble to be saturated, $p_\tu{V}$ can be estimated by quantifying the temperature surrounding the bubble during its collapse. The temperature following the collapse of a laser-induced vapor bubble has been measured by Quinto-Su \etal\cite{10.1038/srep05445}. By fitting a curve to their data, we notice the temperature rise to have a power law dependence on laser energy ($\propto E^{0.5}$). By extrapolating this reported energy range ($E=1-7$ $\upmu$J) to our case ($E=30-60$ $\upmu$J) where $R_\tu{max}$ is within the field of view, we estimate the temperature rise to be $19-27$ $^{\circ}$C, relative to a room temperature of $25^{\circ}$C. We assume this to be the temperature surrounding the bubble during its collapse and determine the corresponding $p_\tu{V}$ based on the saturation pressure of water. In our experiments, we calculate the ratio $t_\tu{col}/R_\tu{max}$ to be $103.52\pm8.6$ ms/m, which is in good agreement with the estimated value of $104.58\pm1.31$ ms/m for an unbound spherical bubble.

\section{Morphologies}
\begin{figure}[htbp]
	\centering
	\includegraphics[width=0.7\columnwidth]{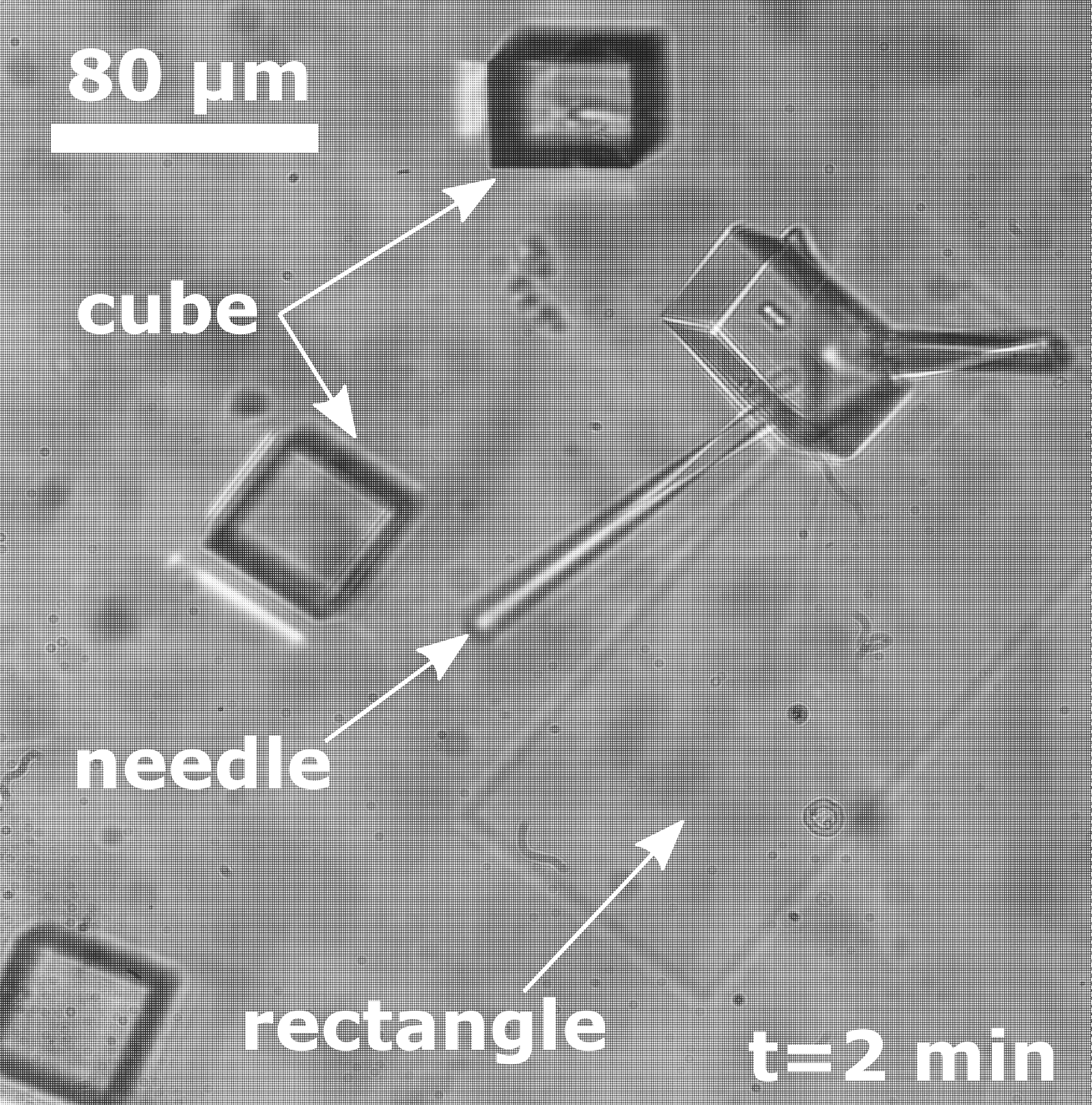}
	\caption{Illustration of \ch{KCl} crystal morphologies observed in experiments.}
    \label{morphology_image}
\end{figure}

We observe three different morphologies of \ch{KCl} in our experiments, represented in Fig. \ref{morphology_image}. The morphologies can be classified based on the number of significant dimensions: cube - three, rectangle - two and needle - one.

\begin{figure}[htbp]
	\centering
	\includegraphics[width=0.9\columnwidth]{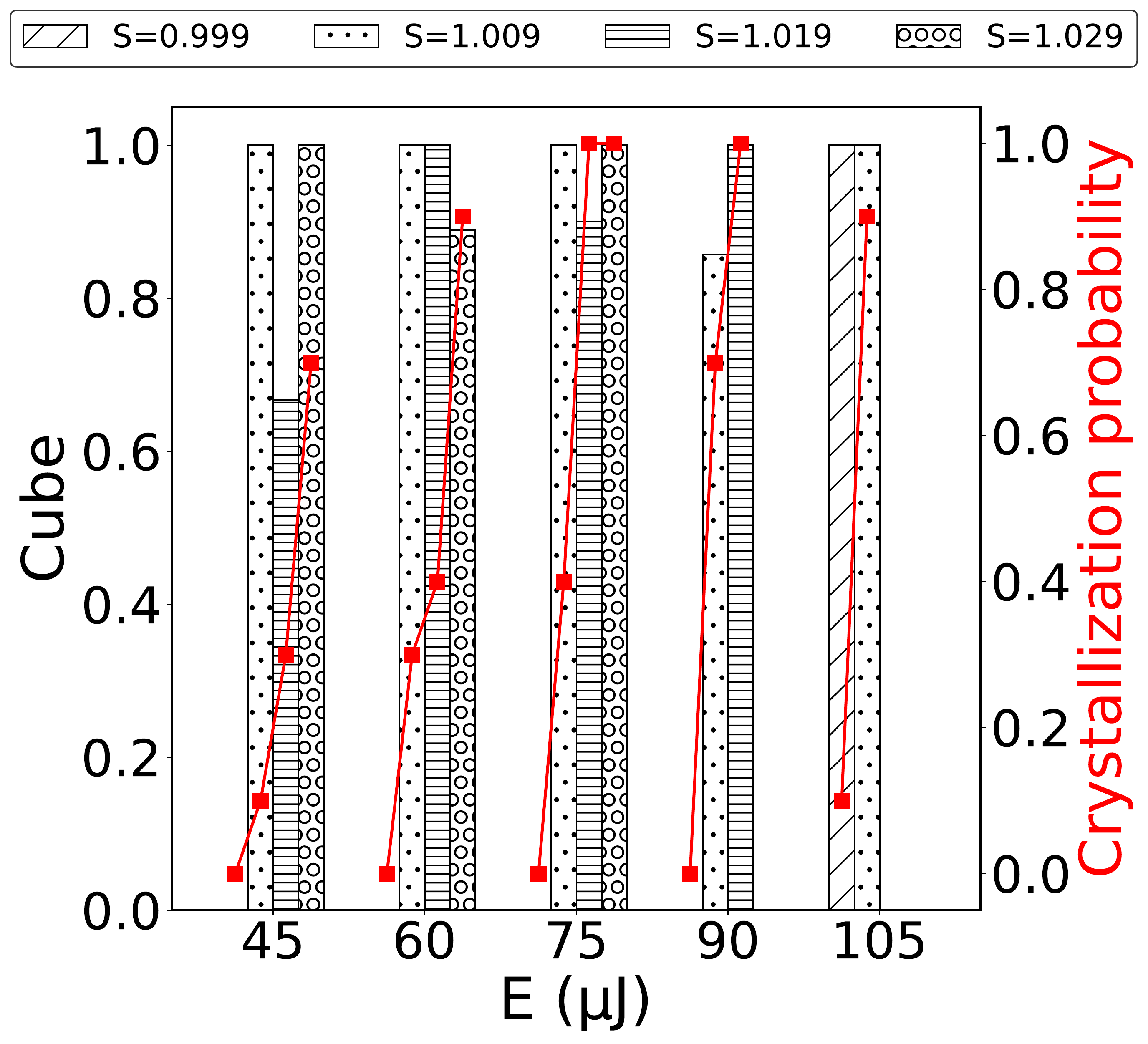}
	\includegraphics[width=0.9\columnwidth]{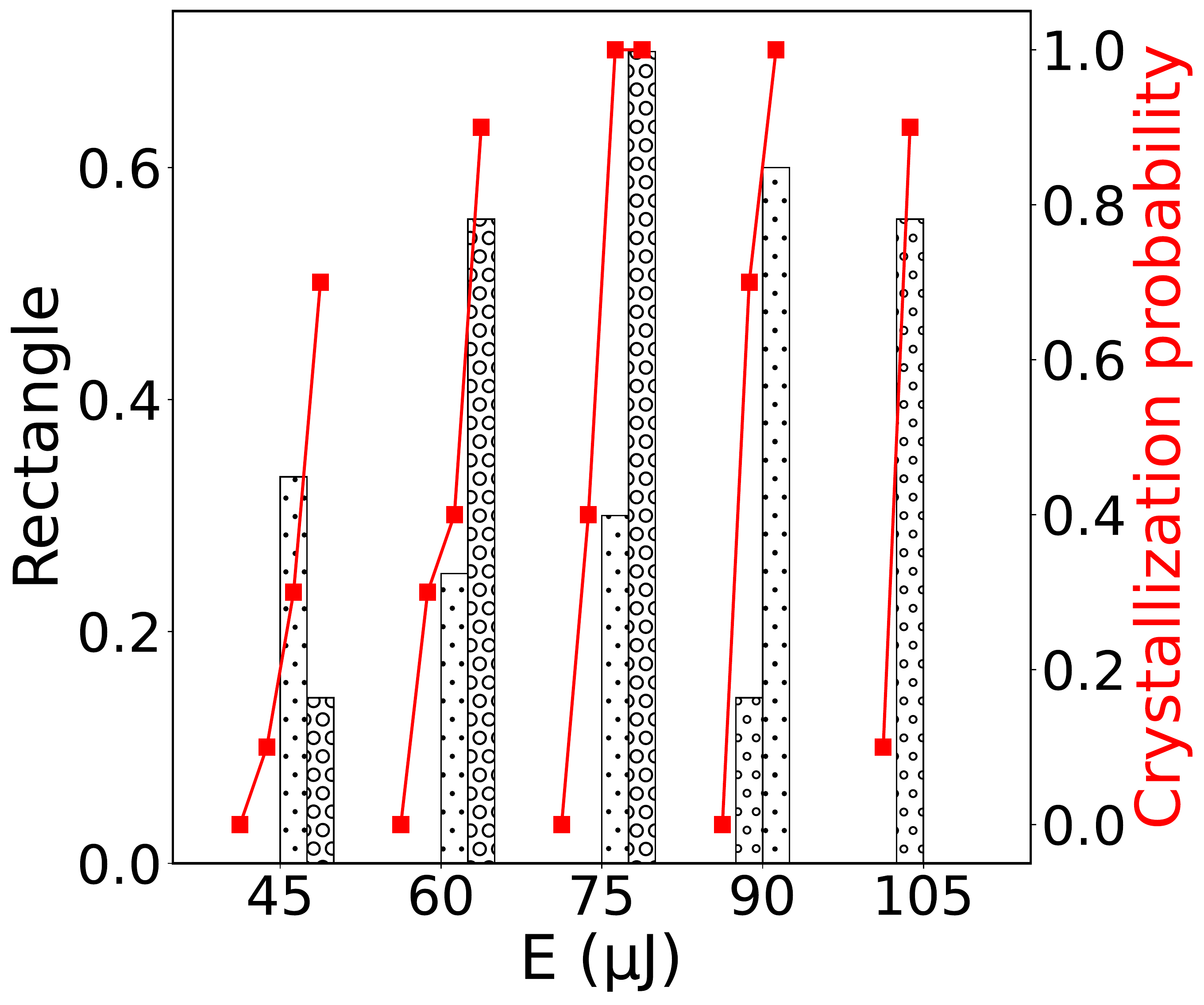}
	\includegraphics[width=0.9\columnwidth]{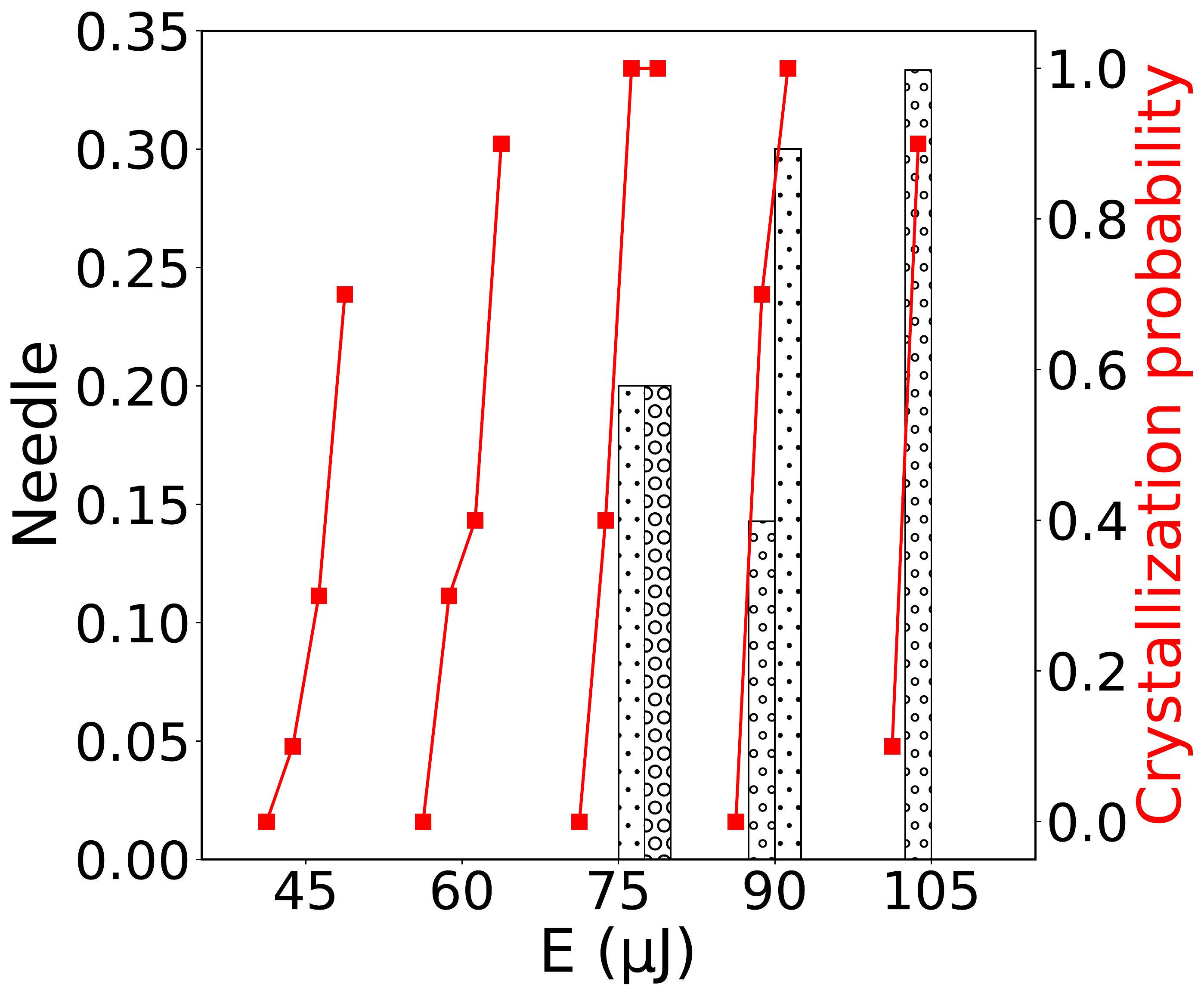}
	\caption{The bars represent the morphology chances from experiments and the red line represents the crystallization probability. $S=S_\tu{bulk}$ -  the supersaturation in the bulk. The crystallization probability is the number of trials that crystallized to the total number of trials performed. The reported data are for a fixed observation time of 2 minutes from the time of the laser pulse.}
    \label{morphology}
\end{figure}

To get better insight on the probability of occurrence of each of the morphology, we calculate the morphology chances. We define the morphology chance as the ratio of number of trials in which we spot a certain morphology to the number of trials that crystallized. Fig \ref{morphology} shows the recorded chances from experiments for each of the morphologies. 

We observe an increase in the probability of finding a rectangle or needle with the increase in bulk supersaturation and laser energy. The increase in the nucleation rate (number of nuclei formed per unit volume per unit time) is non-linear with respect to the local supersaturation (the supersaturation at the vapor-liquid interface in our case) \cite{10.1002/crat.200310070}. Therefore, the amount of solute available in the bulk per nucleus is expected to decrease with increasing laser energies or vapor-liquid interface supersaturation. This explanation is also in agreement with the reported morphological changes based on solute mass transfer \cite{Kardum,10.1021/acs.jpcc.9b11651}.

\section{Simulations: Parameters, Boundary Conditions and Validation}
We approximate the vapor bubble as a lumped system and initialize the model with radius $R(t=0) = 0.5$ $\upmu$m. This initial radius was chosen based on the calculated laser spot size and the objective's depth of field, discussed in Section \ref{Experimental methodology and validation}. At $t=0$, we assume the vapor bubble to be saturated at a temperature of $T_\tu{V}=650$ K - the spinodal temperature of a saturated \ch{NaCl} solution \cite{10.1007/s10953-009-9417-0}. The initial temperature profile of the solution surrounding the vapor bubble is established using $T(\xi) = T_\infty + (T_\tu{V}-T_\infty) \exp[-(\xi / \delta_\tu{T})^{25}]$, where $T_\infty$ is the ambient temperature, $\delta_\tu{T}$ is the thermal boundary layer thickness and $\xi = r-R$, is the radial distance from the interface. A high exponent of 25 is used to approximate a step function while avoiding numerical instabilities near $\xi \approx \delta_\tu{T}$. In this model, we neglect the change in $R$ due to phase change at the vapor-liquid interface. This is justified because we observe a negligible effect of phase change on the liquid velocity, $(\mathrm{d}m_\tu{V}/\mathrm{d}t)/(\rho_\tu{L} A_\tu{V}) \ll \mathrm{d}R/\mathrm{d}t$, in our simulations.

While the momentum equation (Rayleigh–Plesset) is an ordinary differential equation that is solved without the need for meshes, the heat and mass transfer is solved in the frame of reference of the moving vapor-liquid interface. The simulations were performed with COMSOL Multiphysics (version 5.6) using equation-based modeling functionality \cite{COMSOL}. Since all the governing equations under consideration are in spherically symmetric coordinates, we solve them in one dimension.\\

\noindent \textbf{Parameters}\\
\indent $\bullet$ Supersaturation ($S$) and Solubility ($C_\tu{eq}$): solubility of \ch{KCl}, $C_\tu{eq}=35.97$ g/100g-\ch{H2O} at $25^{\circ}$C. Therefore, \ch{KCl} supersaturation in the bulk $S_\tu{bulk}$ = $C$ [concentration in g/100g-\ch{H2O}] / $C_\tu{eq}$ [solubility at $25^{\circ}$C].\\

\indent $\bullet$ Latent heat of vaporization of the solution ($H_\tu{L}$): estimated using the empirical relation $H_\tu{L} = (2230.1-2.80\,C)$ kJ/kg, from the work by Lunnon \cite{10.1088/1478-7814/25/1/317}.\\  

\indent $\bullet$ Thermal conductivity of the solution ($k$): in the work by Ozbek \textit{et al.} \cite{10.2172/6269880}, an empirical correlation equation for aqueous \ch{NaCl} with molality ($m$) 5 mol/kg was reported to be $k = -5.693\times10^{-6}(T-273)^2 + 1.519\times10^{-3}(T-273) + 0.5574$ W/(m\,K). For simplicity, we assume the value to be constant with $k=0.625$ W/(m\,K), since its change between $0-250$ $^{\circ}$C is less than $5\%$.

\indent $\bullet$ Density of the solution ($\rho_\tu{L}$): was measured to be 1173.3 kg/m$^3$ with a density meter (DMA5000, Anton Paar) using a solution with $S=1$. In simulations, the density was assumed to be independent of temperature and supersaturation, with $\rho_\tu{L}=1175$ kg/m$^3$.

\indent $\bullet$ Specific heat capacity of the solution ($c_{p\tu{L}}$): was measured experimentally to be 3000 J/(kg\,K) by Toner \textit{et al.} \cite{10.1021/acs.jced.6b00812} for an aqueous \ch{KCl} solution with molality ($m$) $4.62$ mol/kg.

\indent $\bullet$ Dynamic viscosity of solution ($\mu$): assumed to be constant (0.984 mPa\,s) \cite{10.1063/1.555640} - independent of temperature and pressure. Note that the contribution in the Rayleigh-Plesset equation by the viscosity $(4\mu/R)(\tu{d}R/\tu{d}t)$ is two orders of magnitude less than $(p_\tu{V}-p_{\infty})$ for all conditions studied in this work.

\indent $\bullet$ Surface tension of solution ($\sigma$): assumed to be constant (0.079 N/m) \cite{10.1016/j.mineng.2008.08.001} - independent of temperature and pressure. Note that the contribution in the Rayleigh-Plesset equation by the surface tension $2\sigma/R$ is two orders of magnitude less than $(p_\tu{V}-p_{\infty})$ for all conditions studied in this work.

\indent $\bullet$ Diffusivity of \ch{KCl} ($D$): reported to be $1.92\times 10^{-9}$ m$^2$/s at 25$^{\circ}$C \cite{10.1002/aic.690310603}. Within the solution, we estimate $D$ as a function of temperature to scale according to the Stokes–Einstein equation, $D \propto T/\mu$. The dynamic viscosity as a function of temperature is estimated using the empirical equation from the work by Kestin \textit{et al.} \cite{10.1063/1.555581}.

\indent $\bullet$ Specific heat capacity of vapor ($c_{p\tu{V}}$): the specific heat capacity of superheated steam at high pressure (650\,K and 220\,bar) is calculated to be 2990.81 J/(kg\,K) using the empirical equation by Schmidt \cite{10.2172/4832052}. The specific heat of saturated steam at 50$^{\circ}$C is 1948.24 J/(kg\,K). Since, $H_\tu{L}$ is an order of magnitude higher than $c_{p\tu{V}} (T_\tu{V}-T_\infty)$ in the energy equation for vapor, and in addition, the temperature of the vapor is also governed by the heat convected by the solution surrounding the bubble, we approximate the $c_{p\tu{V}}$ to be constant, 2500 J/(kg\,K).

\indent $\bullet$ Saturation pressure of the solution ($p_\tu{s}$): estimated using the empirical relation $p_\tu{s} = p_\tu{sat}^* (1-0.33\,C/M_\tu{KCl})$, from the work by Leopold \textit{et al.} \cite{10.1021/ja01407a019}. Here $p_\tu{sat}^*$ is the saturation pressure of water, $C$ is the concentration of \ch{KCl} in the solution (in g/100g-\ch{H2O}) and $M_\tu{KCl}$ the molar mass of \ch{KCl}.

\indent $\bullet$ $C^*$ is the concentration of \ch{KCl} in the solution in g/g of solution. $C^* = C/(100+C)$.

\indent $\bullet$ Other constants: specific gas constant, $R_\tu{g}=461.52$ J/(kg\,K). Ambient pressure, $p_{\infty} = 1.013$ bar. Ambient temperature, $T_{\infty} = 25^{\circ}$C. Molar mass of KCl = 74.55 g/mol.\\

\noindent \textbf{Meshing}\\
\indent A finer mesh for the solution domain is established closer to the vapor-liquid interface compared to the farther surroundings. A rough estimate for the penetration depth of solute elements is calculated using $\sqrt{4Dt_\tu{osc}}$, where $t_\tu{osc}$ is the vapor bubble lifetime. Using this, the solute penetration depth at 25$^{\circ}$C for a 100 $\upmu$s bubble is 0.88 $\upmu$m. A uniform grid size of 50 nm is established over a distance of 2 $\upmu$m from the vapor-liquid interface. Then the next 198 $\upmu$m length is divided into 1200 grid cells distributed using an arithmetic sequence with an element ratio of 5 \cite{COMSOL}. The equations are solved using a time step ($\Delta t$) of 1\,ns. With $\alpha$ as the solution's thermal diffusivity and $h$ as the minimum grid size, the maximum value for $\sqrt{4\alpha \Delta t}/h$ and $\sqrt{4D \Delta t}/h$ is 0.53 and 0.2, respectively.

\begin{figure}[htbp]
	\centering
	\includegraphics[width=0.8\columnwidth]{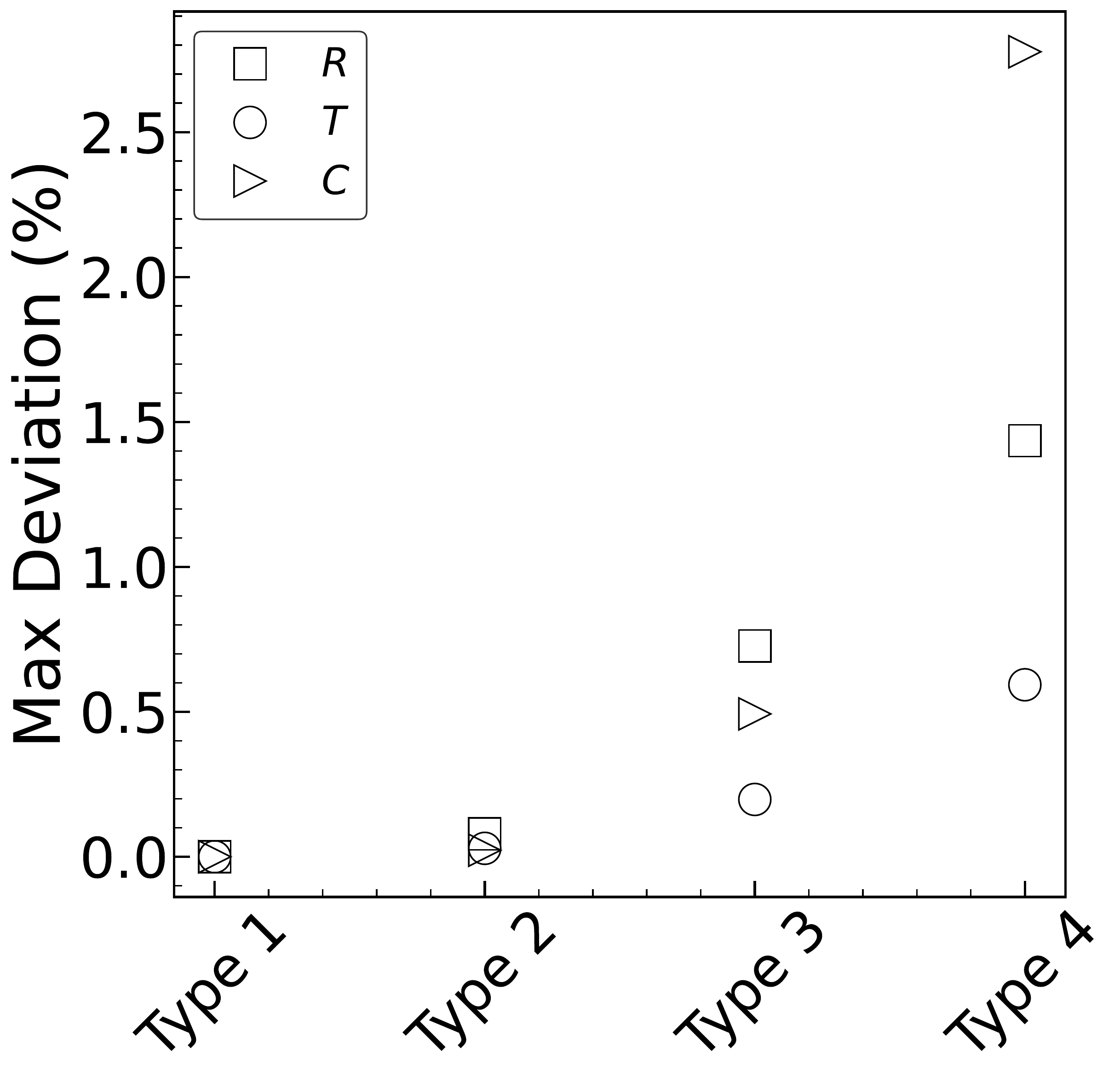}
	\caption{Grid convergence study. The simulations were performed for the grid values mentioned in Table \ref{grid values}. For all the cases, $\sqrt{4\alpha \Delta t}/h \in [0.48,0.6]$ and $\sqrt{4D \Delta t}/h \in [0.16,0.2]$. The maximum deviation percentage was calculated at $t=7\,\upmu$s with respect to the values from Type 1. $R$ is the position of the vapor-liquid interface from the laser focal spot (center), and $T$ and $C$ are the temperature and solute concentration, respectively, in the liquid domain (200 $\upmu$m long) surrounding the vapor bubble. The deviations were calculated with the values from Type 1 as the reference.}
	\label{grid convergence study}
\end{figure}

Figure \ref{grid convergence study} shows the obtained grid convergence for the values presented in Table \ref{grid values}.\\
\begin{table}[htbp]
  \centering
  \caption{Grid values. In all the cases the grid cells inside the domain with 198 $\upmu$m was distributed using an arithmetic sequence with an element ratio of 5 \cite{COMSOL}.}
    \begin{tabular}{>{\centering\arraybackslash}p{3.5em}>{\centering\arraybackslash}p{3em} >{\centering\arraybackslash}p{3em}>{\centering\arraybackslash}p{7em} >{\centering\arraybackslash}p{7em}} \hline
    & $h$ (nm) & $\Delta t$ (ns) & {\footnotesize Number of cells (initial 2 $\upmu$m)} & {\footnotesize Number of cells (farther 198 $\upmu$m)}\\ \hline
    \textbf{Type 1} & 12.5 & 0.05 & 160 & 4800\\ 
    \textbf{Type 2} & 50 & 1 & 40 & 1200\\ 
    \textbf{Type 3} & 200 & 20 & 10 & 300\\ 
    \textbf{Type 4} & 500 & 100 & 4 & 120\\\hline 
    \end{tabular}%
    \label{grid values}
\end{table}%

\noindent \textbf{Boundary conditions}\\
\indent $\bullet$ For the heat transfer within the solution, we apply Dirichlet conditions at either end, with $T=T_V$ at the vapor-liquid interface and $T=T_\infty$ at the exterior domain of the simulation.

\indent $\bullet$ For the mass transfer within the solution, we apply a Neumann condition at the vapor-liquid interface, with the solute flux given by $(\tu{d}m_\tu{V}/\tu{d}t) C_\tu{bulk}/ (\rho_\tu{L} A_\tu{V})$, and a Dirichlet condition $C=C_\tu{bulk}$ at the exterior domain of the simulation.

\begin{figure}[htbp]
\centering
\includegraphics[width=0.75\columnwidth]{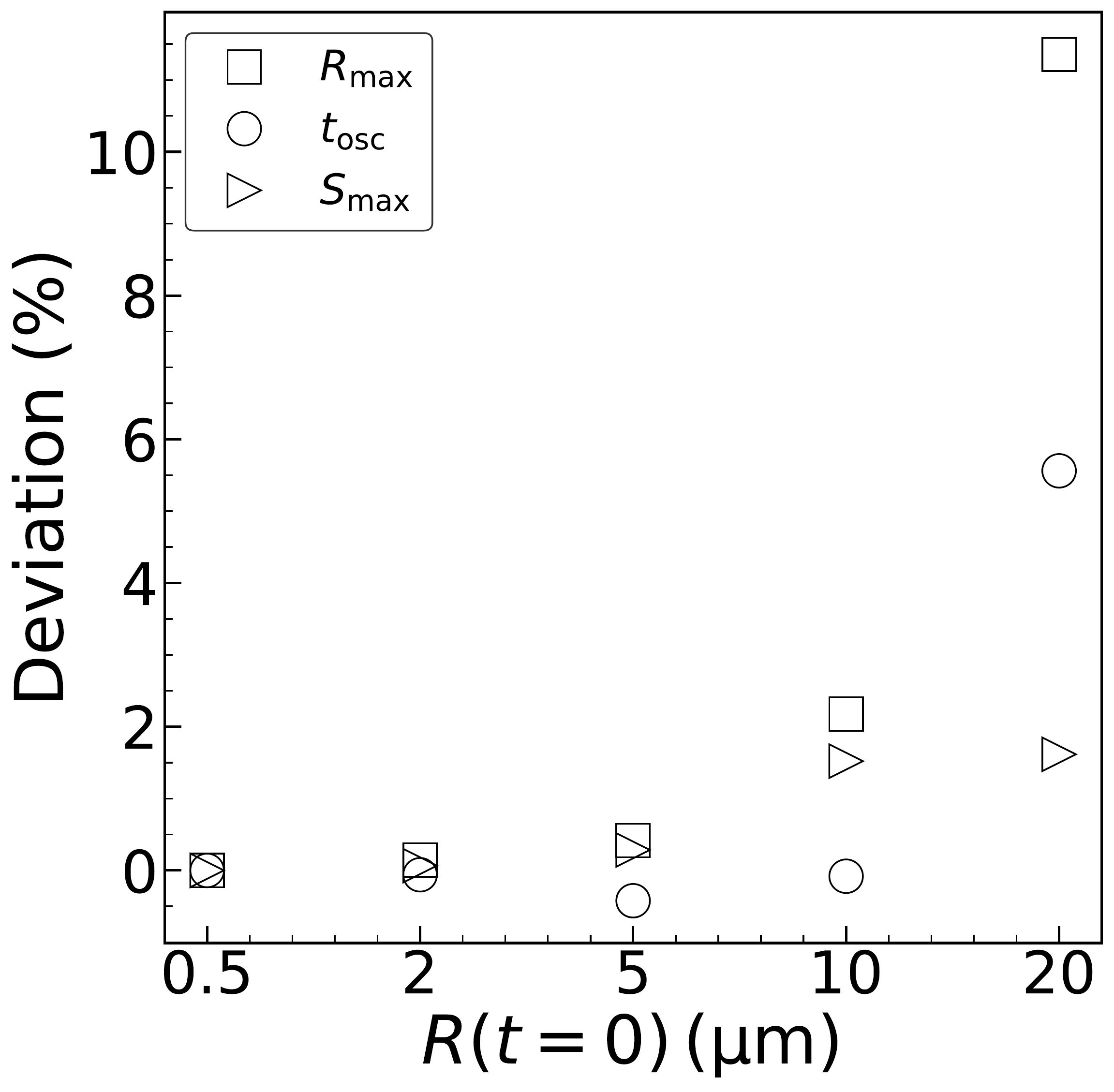}
\caption{Sensitivity analysis for different initial radii of the bubble in simulations. The deviation is calculated using the reference values from $R(t=0)=0.5\,\upmu$m ($R_\tu{max}=228\,\upmu$m). The simulations are for $E=43.14\,\upmu$J and $S_\mathrm{bulk}=1.019$ (at $25^{\circ}$C). The value of $\delta_\mathrm{T}(\upmu\tu{m})\in\{25.5,24,21,16.3,8.35\}$ correspond to $R(t=0)\in\{0.5,2,5,10,20\}$, respectively.}
\label{sensitivity analysis}
\end{figure}

\indent $\bullet$ Sensitivity analysis of the key parameters, such as: $R_\tu{max}$, $t_\tu{osc}$ and $S_\tu{max}$ for different $R(t=0)$ was performed. We observe a small effect on the parameters ($<12\%$), even when $R(t=0)$ is increased by a factor of 40, Fig. \ref{sensitivity analysis}. The analyses were performed by fixing the $E=43.14\,\upmu$J and $S_\mathrm{bulk}=1.019$ (at $25^{\circ}$C). The value of $\delta_\mathrm{T}$ is changed with $R(t=0)$ to maintain the $E$ constant. This demonstrates the efficiency of the model and validity of the boundary conditions. Especially, it emphasizes the only tunable parameter in simulations, $\delta_\mathrm{T}$, as a parameter that characterizes the energy surrounding the initial bubble (similar to laser energy in experiments) and is not explicitly a fudge factor.\\

\noindent \textbf{Energy parameter $\delta_\tu{T}$}

$\delta_\mathrm{T}$ characterizes the initial thermal energy surrounding the bubble to be used for growth in the simulations. Since we only match the dynamic bubble radius from experiments to simulations, we indirectly compare the energy supplied by the laser against the $\delta_\mathrm{T}$ from simulations. Analytically the energy supplied in the simulations can be calculated as,
\begin{align}
\begin{split}
    E_\mathrm{sim} &= m_\tu{V} (H_\tu{L} + c_{p\mathrm{V}} (T_\tu{V}-T_\infty)) \,+ 4\pi \rho_\tu{L} c_{p\tu{L}}\\
    & \int^\infty_{0} (R(t=0)+r)^2 (T_\tu{V}-T_\infty) \exp[-(r / \delta_\tu{T})^{25}] \,\tu{d}r.
    \label{EEq}
\end{split}
\end{align}
The values for the variables in Eq.\,\ref{EEq} are chosen at $t=0$, and therefore excludes the kinetic energy. Energy surrounding a hemisphere is $E = E_\mathrm{sim}/2$. The dependence of $R_\mathrm{max}$ on $E$ from simulations is represented in Fig.\,5 of the main text.

In our experiments, since the bubble momentarily goes out of the field of view for higher energies ($E\ge75\,\upmu$J), for all cases, we numerically simulate the experiments using the bubble's lifetime ($t_\mathrm{osc}$). This approach to plotting experiments against simulations is rationalized using the Rayleigh collapse time equation, $t_\mathrm{osc}/R_\mathrm{max} = 1.83 \left[\rho_\tu{L}/(p_\infty-p_\mathrm{V})\right]^{1/2}$, which relates the bubble lifetime ($t_\mathrm{osc}$) and maximum bubble size ($R_\mathrm{max}$). As the initial thermal energy surrounding the bubble is what dictates its dynamics, we determine the numerical energy value from the model. Since we have fixed the initial bubble size ($R=0.5\,\upmu$m) and thermal boundary layer profile ($T(\xi) = T_\infty + (T_\mathrm{V}-T_\infty) \exp[-(\xi / \delta_\mathrm{T})^{25}]$) in the simulations, we can calculate the value of $\delta_\mathrm{T}$ using the numerical thermal energy (Eq.\,\ref{EEq}).

For the calculations and validation of the theory for Rayleigh collapse time against the experiments performed in this work, refer to Section \ref{Experimental methodology and validation}.\\

\begin{figure}[htbp]
\centering
\includegraphics[width=\columnwidth]{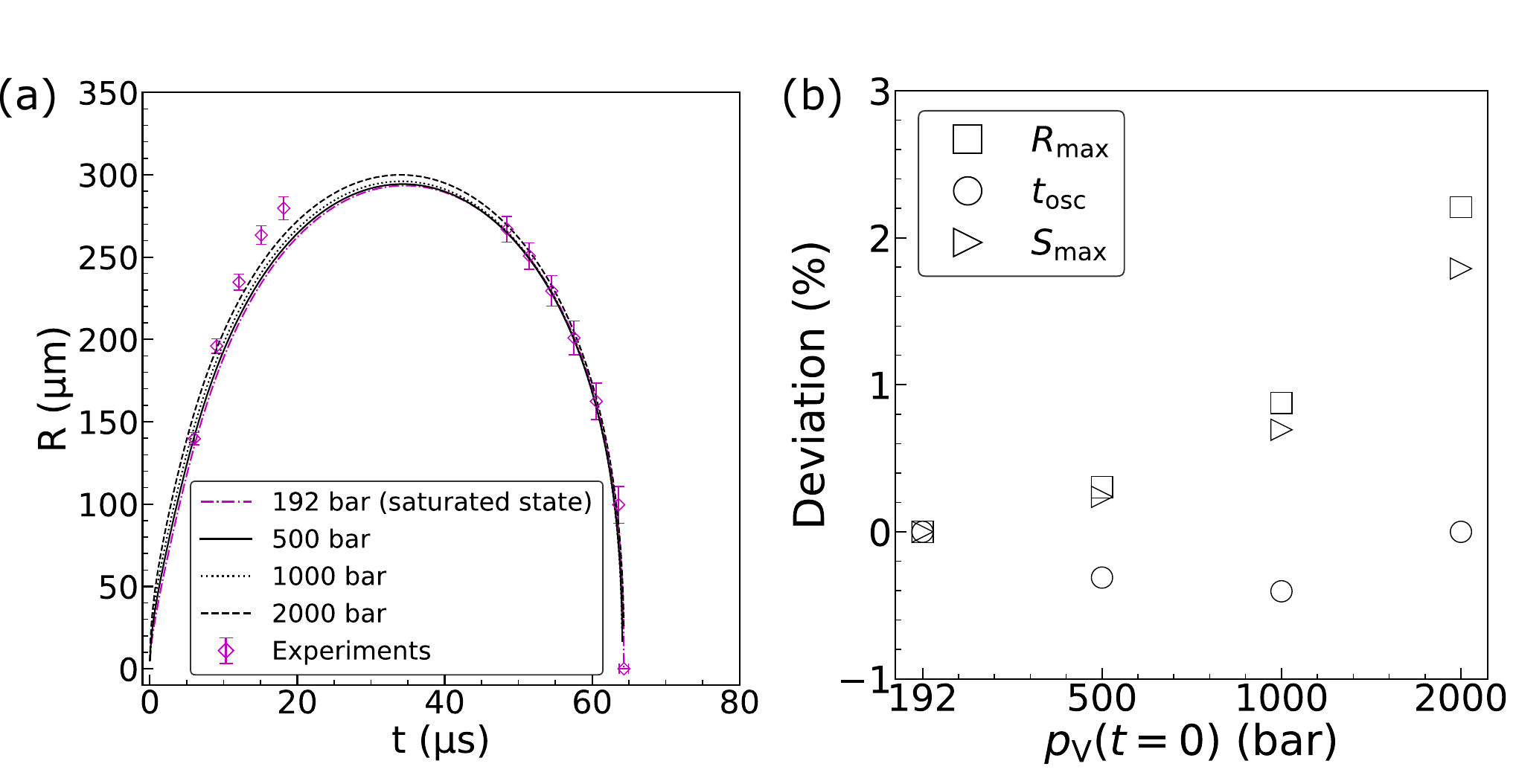}
\caption{(a) Sensitivity analysis for different initial pressure of the bubble in simulations. The experiment is for $E=75\,\upmu$J. The simulations are for $\delta_\mathrm{T}=28\,\upmu$m with $R(t=0)=5\,\upmu$m. (b) The deviation is calculated using the reference values from 192\,bar.}
\label{pV}
\end{figure}

\noindent \textbf{Effect of initial bubble pressure}\\
\indent While the thermal energy surrounding the bubble influences the $R_\tu{max}$, the initial bubble pressure would influence $\tu{d}R/\tu{d}t$ close to formation. In simulations, violating our current assumption of a saturated bubble at $t=0$, we intentionally supply the bubble with very high pressure of $2000$\,bar. Thus, while saturated vapor bubble in a supersaturated aqueous solution has 192\,bar @ 650\,K \cite{10.1007/s10953-009-9417-0}, the condition is changed to 500-2000\,bar @ 650\,K. The maximum temperature of the liquid surrounding the bubble was maintained 650\,K - the spinodal temperature limit \cite{10.1007/s10953-009-9417-0}. In addition, the initial bubble radius, $R(\tu{t=0})$, was changed from $0.5\upmu$m to $5\upmu$m to increase the mechanical energy ($E_\tu{bubble} \propto p_\tu{V}\,V_\tu{V}$) \cite{10.1103/PhysRevLett.88.078103}. The model is less sensitive for $R(t=0)=0.5-5\,\upmu$m (refer Fig.\,\ref{sensitivity analysis}), provided we supply the same thermal energy surrounding the bubble. The results are presented in Fig.\,\ref{pV}.

As we increase the bubble pressure at $t=0$ we notice a decrease in deviation in dynamic bubble size ($R$) between the experiments and simulation. Such high initial bubble pressures [$O(1000\,\tu{bar})$] can exist only if non-linear absorption is involved with potential plasma formation \cite{10.1121/1.415878}. The argument is in agreement with our experimental observation on the increase in probability of bubble formation for $E\ge 75\,\upmu$J in absence of \ch{KMnO4}, reported in Section \ref{Bubble incidence in experiments}. As the $\tu{d}R/\tu{d}t$ increases the rate at which the pressure drops within the bubble increases. Consequently, the difference between the bubble pressure and the saturation pressure of the liquid at the interface increases leading to increased solvent evaporation rates, estimated using Hertz-Knudsen equation \cite{10.1016/0021-9797(92)90205-Z}. Thus the increase in initial vapor pressure reduces the mismatch in $R$ between experiments and simulations, with a consequent increase in the interface peak supersaturation ($S_\tu{max}$). The observed increase in $S_\tu{max}$ further validates the proposed nucleation hypothesis based on solute accumulation at the interface.\\

\begin{figure}[htbp]
	\centering
	\includegraphics[width=0.8\columnwidth]{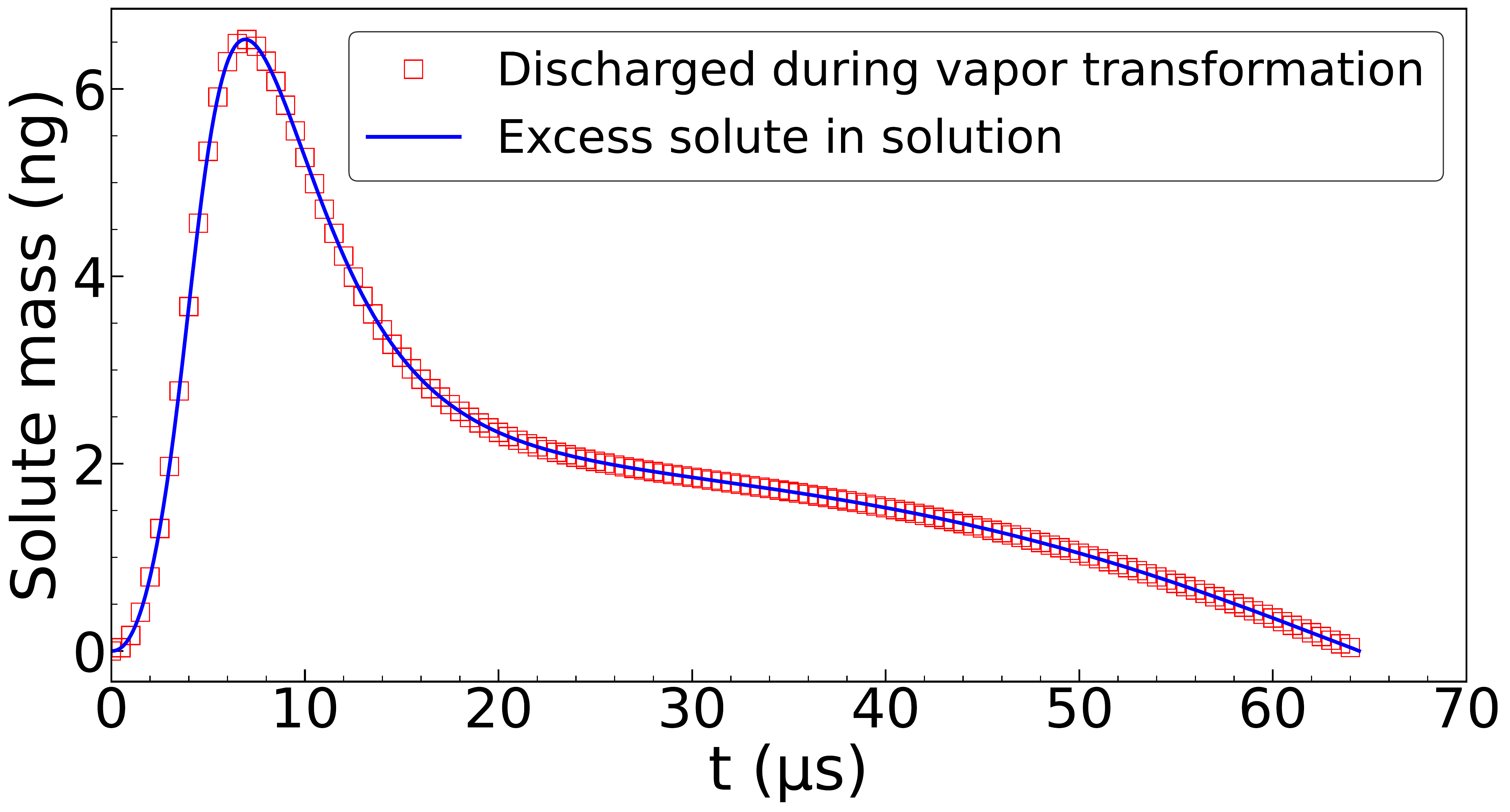}
	\caption{Solute conservation. The represented data from simulations is for $\delta_\tu{T} = 32.5$ $\upmu$m ($E=88.2$ $\upmu$J - for a hemispherical vapor bubble) and $S_\tu{bulk} = 1.019$.}
	\label{solute conservation}
\end{figure}

\begin{figure}[htbp]
	\centering
	\includegraphics[width=0.8\columnwidth]{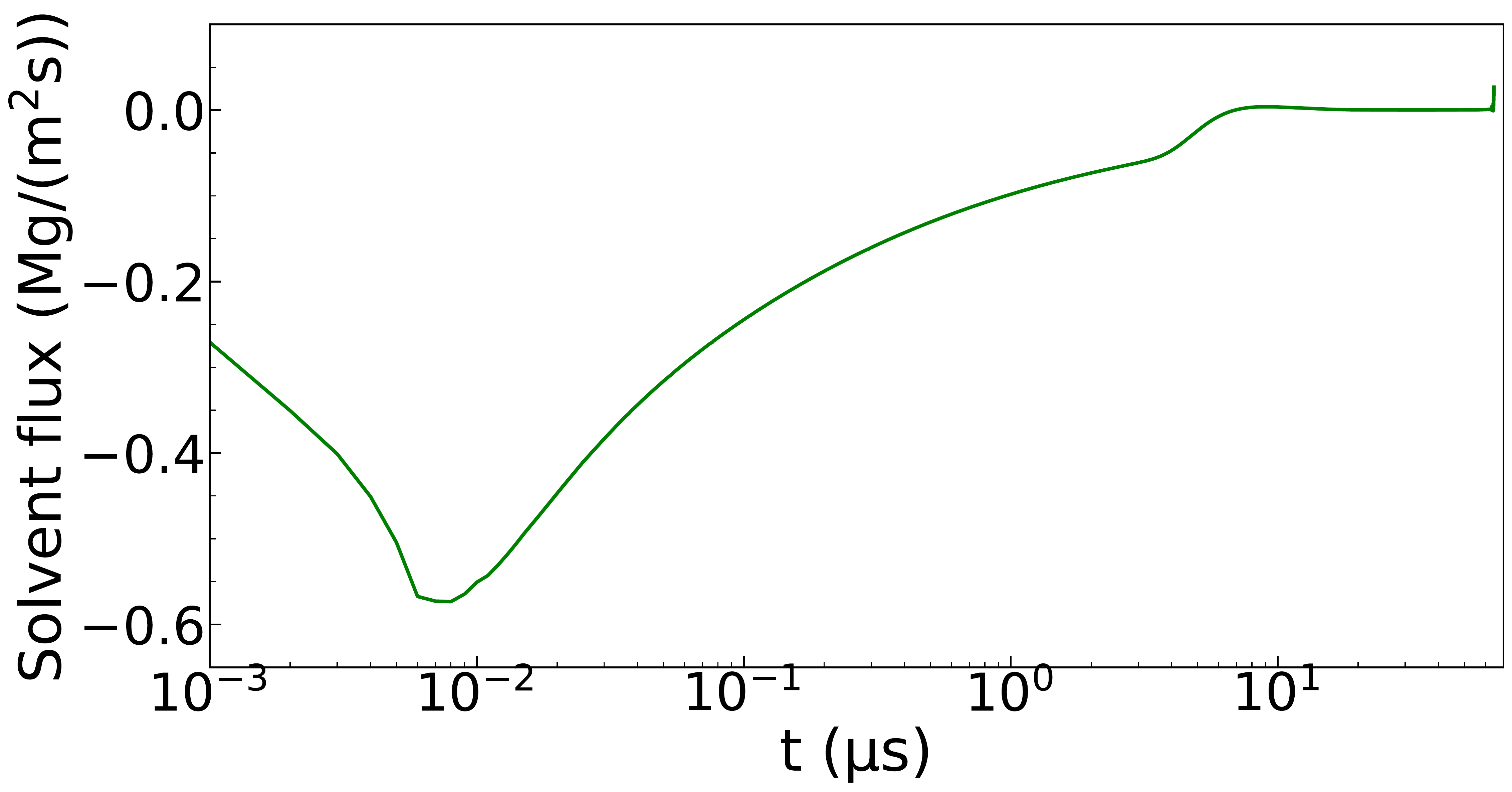}
	\caption{Solvent mass flux density at the vapor-liquid interface. The negative sign indicates evaporation. The represented data from simulations is for $\delta_\tu{T} = 32.5$ $\upmu$m ($E=88.2$ $\upmu$J - for a hemispherical vapor bubble) and $S_\tu{bulk} = 1.019$.}
	\label{mass flux density}
\end{figure}

\noindent \textbf{Solute and heat conservation}\\
\indent The conservation of solute mass is shown in Fig \ref{solute conservation}. When the solution evaporates, the dissolved solute is assumed to remain in the liquid phase. Therefore, at any time instant, the solute mass that was dissolved in the evaporated solvent ($m_\tu{V}C_\tu{bulk}$) should be equal to the excess solute present in the solution ($\int_R^\infty 4 \pi r^2 \rho_\tu{L} (C^*-C_\tu{bulk}^*) \tu{d}r$). Fig \ref{mass flux density} shows the corresponding mass flux density of the solvent at the vapor-liquid interface.

\begin{figure}[htbp]
	\centering
	\includegraphics[width=0.8\columnwidth]{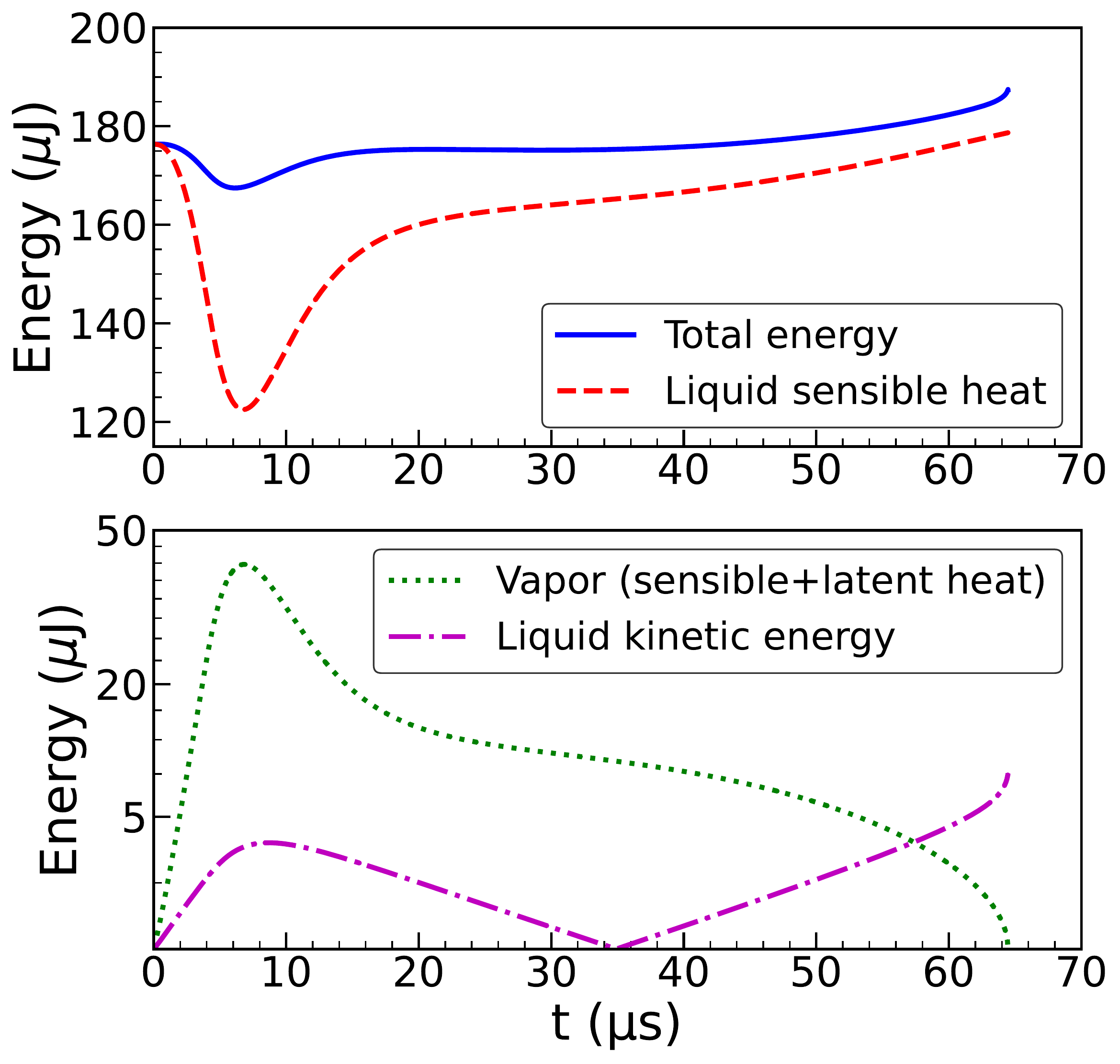}
	\caption{Energy trends. The represented data from simulations is for $\delta_\tu{T} = 32.5$ $\upmu$m ($E=88.2$ $\upmu$J - for a hemispherical vapor bubble).}
	\label{energy trends}
\end{figure}
Due to the action of non-conservative force as a consequence of solution viscosity ($\propto 4\mu \Dot{R}/(\rho_\tu{L} R)$), the total energy is not conserved. Moreover, the force due to surface tension ($\propto 2\sigma/(\rho_\tu{L} R)$) and external force due to ambient pressure ($p_\infty$) are expected to do negative work on the system during bubble expansion and positive work during the collapse, Fig. \ref{energy trends}. The total energy equals the sum of sensible heat (vapor and liquid), latent heat (vapor) and kinetic energy of the solution at any time instant. At the start of the process, the thermal energy of the liquid decreases due to evaporation. Correspondingly, we see an increase in the energy content of the vapor and the kinetic energy of the solution. As expected, these trends reverse in magnitude during condensation. 

\section{Solute-solution interfacial tension calculation}
The rate of nucleation (the number of nuclei formed per unit time per unit volume) is expressed using \cite{10.1002/crat.200310070}, 
\begin{equation}
    J \propto S\exp[-16\pi v^2\gamma^3 / (3k_\tu{B}^3 T^3\log^2(S))].
    \label{J eqn}
\end{equation}
The following table provides the values used for calculation.
\begin{table}[htbp]
  \centering
  \caption{Physical and derived parameter values for aqueous \ch{KCl} solution.}
    \begin{tabular}{>{\centering\arraybackslash}p{4.5em}>{\centering\arraybackslash}p{12.5em} >{\centering\arraybackslash}p{8.5em}} \hline
    \textbf{Parameter} & \textbf{Description} & \textbf{Value} \\ \hline
    $v$ & Solute molecular volume & $6.238\times10^{-29}$ m$^3$\\
    $k_\tu{B}$ & Boltzmann constant & $1.381\times10^{-23}$ m$^2$kg/(s$^2$K)\\ 
    $B$ & Gradient(slope) for $\log(J/S)$ vs 1/$\log^2(S)$ [Eq. (\ref{J eqn})] & $6.02\pm3.95$\\
    $\rho_\tu{KCl}$ & Solute density & 1989 kg/m$^3$ \\
    $\rho_\tu{L}$ & Solution density (at 25\,$^{\circ}$C) & 1175 kg/m$^3$ \\
    $\rho_\tu{L}$ & Solution density. Obtained by curve fit using the data from Ref. \cite{10.3133/ofr84253} ($T$ in K) & $(-4.41/(48.28+T-273))+1.24$ g/m$^3$ \\
    $M_\tu{KCl}$ & Molar mass of solute & 74.55 g/mol \\
    $N_\tu{A}$ & Avogadro constant & $6.023\times10^{23}$ mol$^{-1}$\\
    \hline
    \end{tabular}%
    \label{physical and derived parameters}
\end{table}%

From the numerical simulations performed, the maximum supersaturation ($S_\tu{max}$) was found to occur between $185-191\,^{\circ}$C. Thus, the $\gamma$ from Eq. (\ref{J eqn}) was calculated to be $3.7\substack{+0.47 \\ -0.65}$ mJ/m$^2$ using the average temperature of 461\,K and the slope value of $12.75\pm5.53 \times 10^{-3}$ obtained from Figure 5(a) in the main text.

Mersmann calculated a simple relation for the solute-solution interfacial tension based on thermodynamics as \cite{10.1016/0022-0248(90)90850-K},
\begin{equation}
    \gamma \propto T \log\left({\frac{\rho_\tu{KCl}}{C^*\,\rho_\tu{L}}}\right).
    \label{gamma}
\end{equation}
The values for the parameters are taken from Table \ref{physical and derived parameters}. $C^*$ is the solute concentration in kg/kg of solution. During the event when maximum supersaturation at the vapor-liquid interface occurs: using the values of $C^*=0.46$\,kg/kg and $\rho_\tu{L}=1221$\,kg/m$^3$ at 461\,K, we calculate $\gamma=3.51$ mJ/m$^2$ for 298 K.

\begin{figure}[htbp]
	\centering
	\includegraphics[width=0.8\columnwidth]{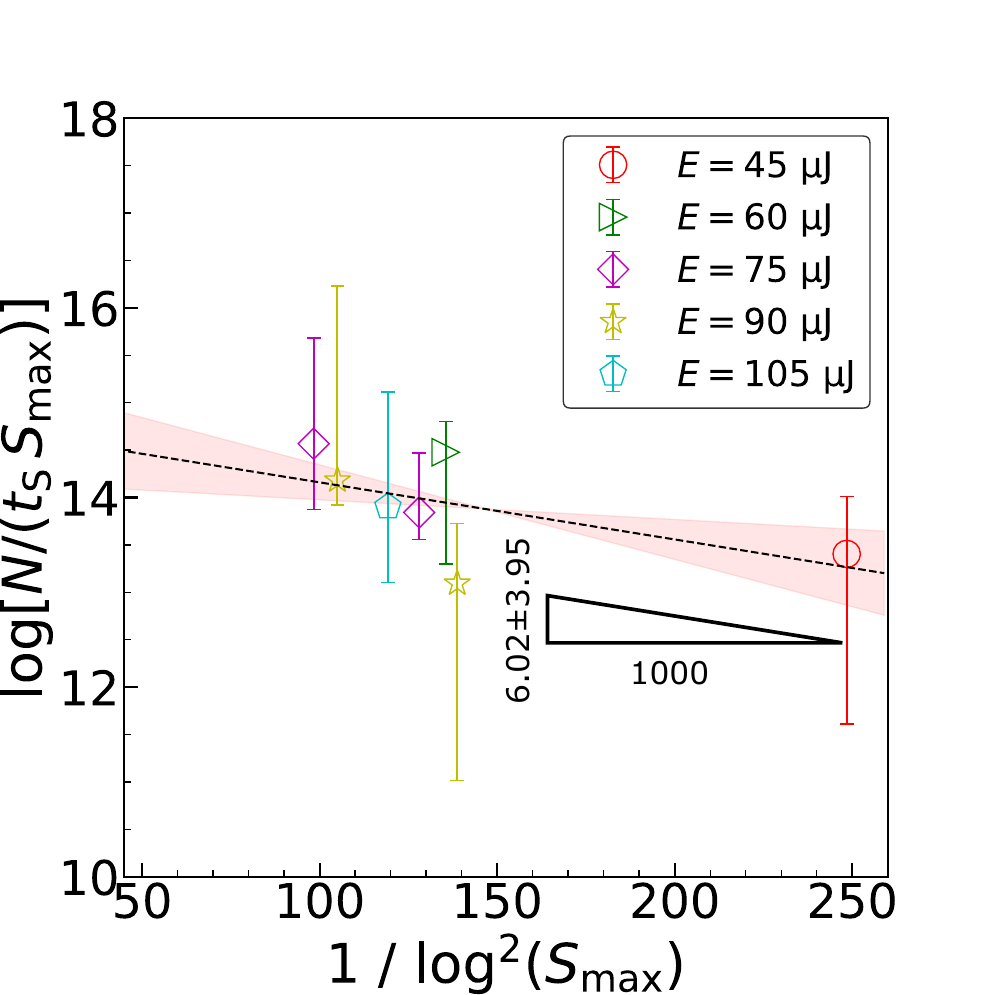}
	\caption{Relation between the experimentally observed median of crystal count ($N$ is the median only in this case) and simulated peak supersaturation ($S_\tu{max}$). $N/t_\tu{s} \propto J$, where $t_\tu{s}$ is the time for which $S>1$ in simulations and $J$ the nucleation rate. The error bars represent the interquartile ranges.}
	\label{J_median}
\end{figure}
Figure \ref{J_median} shows the slope obtained when the median of the crystal count observed in experiments is taken into account for calculation. By using the value of this slope in Eq. (\ref{J eqn}), we estimate the $\gamma$ value to be $2.88\substack{+0.53 \\ -0.86}$\,mJ/m$^2$ (at $\approx 185-191 \,^{\circ}$C). This obtained value, when calculated for $25\, ^{\circ}$C (2.73\,mJ/m$^2$), is in excellent agreement with the $\gamma$ values reported for supersaturated \ch{KCl} solutions, $2.73-2.91$\,mJ/m$^2$ \cite{10.1021/ie50510a038,10.1021/j150569a014,10.22034/ijnc.2019.33448}. From the analysis, we observe the median value of the crystal count from experiments to closely estimate the $\gamma$ values reported for supersaturated \ch{KCl} solutions, compared to mean values. The reasoning lies in the statistical approach since, in this work, we have performed 10 trials for each data point and the recorded data is skewed. Therefore median is a good measure of central tendency.        


\section{Miscellaneous Calculations}

\noindent \textbf{Estimate of the $R_\tu{max}$ in experiments}\\
\indent In literature, the lifetime of the bubble ($t_\tu{osc}$ - time taken for one oscillation) is roughly estimated as \cite{10.1017/jfm.2015.183,10.1529/biophysj.105.079921}, $t_\tu{osc} = 2\,t_\tu{col}$. Thus from Eq.\,(\ref{t_osc}), we can estimate the $R_\tu{max}$ in our experiments using $t_\tu{osc}$. The $p_\tu{V}$ in the equation is estimated assuming saturated conditions where the temperature is obtained by extrapolating the values for the reported energy range ($E=1-7$ $\upmu$J) by Quinto-Su \etal\cite{10.1038/srep05445}.\\  

\noindent \textbf{Shock waves}\\
\indent Explosive formation of a cavitation bubble or its implosion (collapse) can lead to transient high-pressure environment (shockwaves, $\sim$GW/cm$^2$) surrounding the bubble \cite{10.1103/PhysRevLett.88.078103,10.1103/PhysRevLett.73.2853}. These emitted shock waves would compress the liquid as it propagates outwards and decays over time. We determine the significance of shock wave in crystal nucleus formation to eliminate its possibility and further strengthen the mechanism based on evaporation leading to supersaturation spike at the vapor-liquid interface.  

The change in chemical potential $(\Delta \mu)_{\Delta p}$ of the solution due to pressure changes is quantitatively estimated using \cite{10.1021/acs.cgd.6b01437},
\begin{equation}
    (\Delta \mu)_{\Delta p} = \Delta V_\tu{m} \Delta p - \Delta S_{m} \Delta T.
\end{equation}
Where $\Delta V_\tu{m}$ and $\Delta S_\tu{m}$ are the change in the volume and entropy per mole of the solute. Conditions under which $(\Delta \mu)_{\Delta p} > 0$ should result in an increase in crystallization probability due to the shock wave. For simplicity, we assume the solution properties to be that of water and use the equation of state of water to calculate the changes \cite{10.1063/1.1743415}, refer Table \ref{equation of state of water}. For example, the parameter values calculated for $\Delta p = 1\,\tu{GPa}$ are $\Delta V_\tu{m} = 37.02\,\upmu$m$^3$/mole, $\Delta S_\tu{m} = 15.61$ J/(K mole) and $\Delta T = 36$ K.

Venugopalan \etal\cite{10.1103/PhysRevLett.88.078103} measured the size and shape of the laser-induced plasma together with the pressure of the emitted shock wave. In Section \ref{Experimental methodology and validation}, we have theoretically calculated the diameter and length of the laser spot to be 0.762 $\upmu$m and 1.71 $\upmu$m, respectively. For simplicity, we assume the plasma to be spherical and take the initial plasma diameter to be 1.236 $\upmu$m. Since the experimentally measured plasma volume was found to be larger than the theoretical volume by a factor of $\approx16$ \cite{10.1103/PhysRevLett.88.078103}, in our calculations, we correct the initial plasma diameter to 4 $\upmu$m.

\begin{figure}[htbp]
	\centering
	\includegraphics[width=0.8\columnwidth]{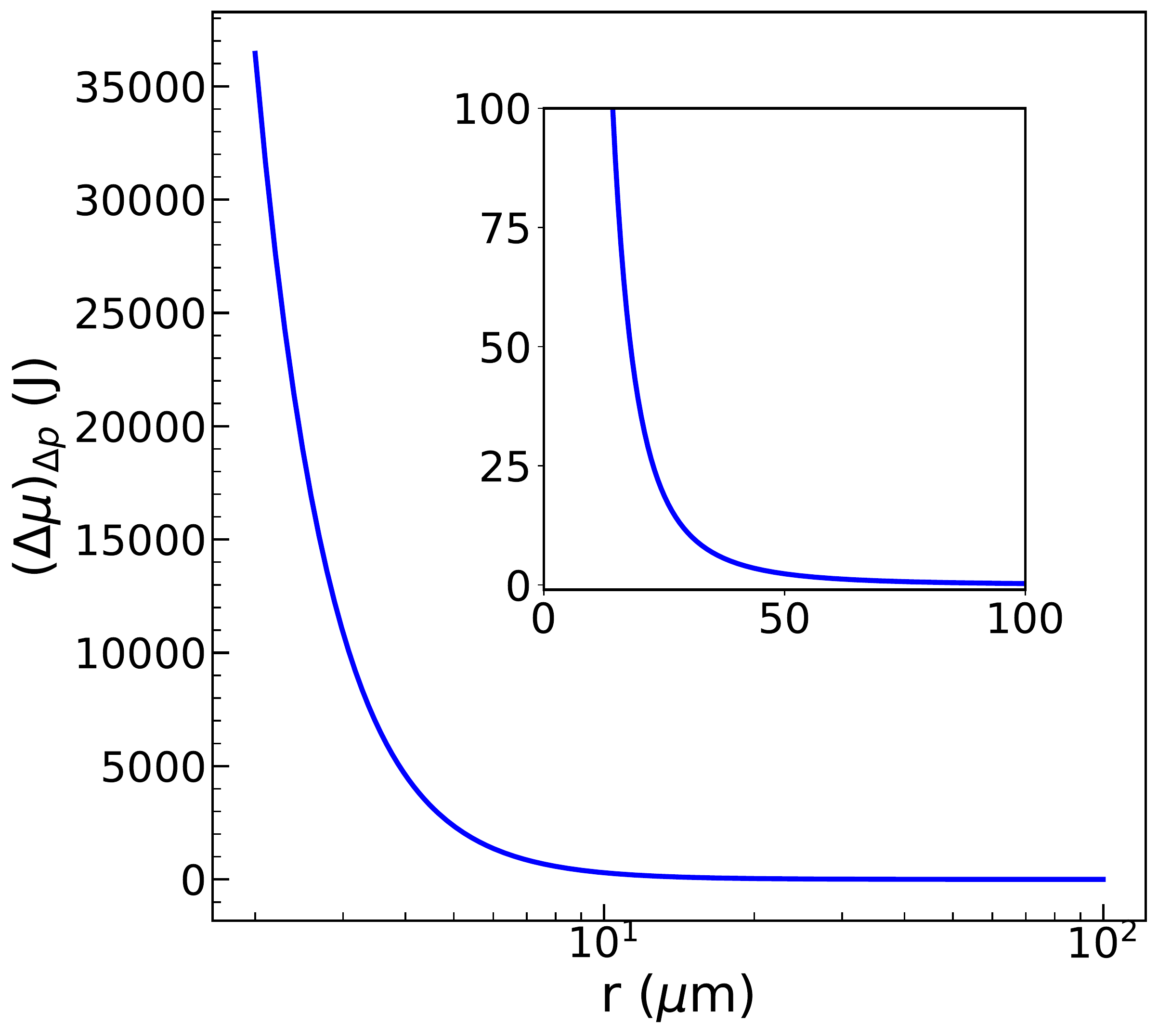}
	\caption{The change in solution's chemical potential due to pressure ($(\Delta \mu)_{\Delta p}$) at different positions from the laser focal spot (as the shock wave propagates). The curve starts at $r=2\, \upmu$m, the initial radius of the speculated plasma.}
	\label{delta_f}
\end{figure}
Figure \ref{delta_f} shows the calculated value of $(\Delta \mu)_{\Delta p}$ at different radial positions from the laser focal spot (bubble center). In this calculation, the pressure was assumed to radially decay with $r^{-2}$, with the pressure at the vapor-liquid / plasma-liquid interface to be 1 GPa \cite{10.1103/PhysRevLett.88.078103,10.1121/1.415878}. The $(\Delta \mu)_{\Delta p}$ is found to sharply decrease in the vicinity of plasma/vapor bubble ($\lesssim 40\,\upmu$m). Although $(\Delta \mu)_{\Delta p} > 0$ favors the formation of a crystal nucleus, the length scale is comparable to that of the thermal boundary layer thickness ($\delta_\tu{T}$). Therefore, we expect no formation of crystals
due to shockwaves because of the lower supersaturation ratio associated with higher temperatures. From Fig. \ref{plasma incidence}, the probability of plasma formation is significant close to $90\, \upmu$J, which corresponds to $\delta_\tu{T}$ value of $\approx 32.5\,\upmu$J (Figs. \ref{solute conservation} and \ref{mass flux density}). As a result, although pressure waves might favor the crystal nucleus formation, the absorbed energy by the liquid from the incident laser would decrease the supersaturation and hence the nucleation probability. Recent evidence on the difference in crystal size distribution between shock wave and direct laser-induced crystallization \cite{10.1021/acsomega.2c03456} also suggests a mechanism other than shock wave is at play.   

\begin{table}[htbp]
  \centering
  \caption{Equation of state of water from shock wave measurements. The equations for $\Delta S_\tu{m}$, $\Delta V_\tu{m}$ and $\Delta T$ are obtained by curve fitting the values from Rice \etal\cite{10.1063/1.1743415} for $\Delta p\in\{0,1.5\,\tu{GPa}\}$. The $\Delta p$ in the equations are in units of GPa.}
    \begin{tabular}{>{\centering\arraybackslash}p{4.5em}>{\centering\arraybackslash}p{11em} >{\centering\arraybackslash}p{10em}} \hline
    \textbf{Parameter} & \textbf{Description} & \textbf{Value} \\ \hline
    $\Delta V_\tu{m}$ & Change in molar volume & $0.1786\, M_\tu{KCl} \Delta p^{0.5019} / C$ [$\upmu$m$^3$/mole]\\
    $\Delta S_\tu{m}$ & Change in molar entropy & $0.0753\, M_\tu{KCl} \Delta p^{1.844} / C$ [J/(K mole)]\\ 
    $\Delta T$ & Change in temperature & $36\,\Delta p$ [K]\\
    \hline
    \end{tabular}%
    \label{equation of state of water}
\end{table}%

\newpage
\bibliographystyle{apsrev4-2}
\bibliography{refs}